\documentclass[final,5p,times,twocolumn]{elsarticle}

\def\BibTeX{{\rm B\kern-.05em{\sc i\kern-.025em b}\kern-.08em
    T\kern-.1667em\lower.7ex\hbox{E}\kern-.125emX}}

\usepackage{amsmath,amssymb,amsfonts}
\usepackage{algpseudocode}
\usepackage{algorithm}

\usepackage{graphicx}
\usepackage{subfig}
\usepackage{textcomp}
\usepackage[table,xcdraw]{xcolor}
\usepackage{url}
\usepackage{multirow}
\usepackage{booktabs}
\usepackage{balance}
\usepackage{caption}
\usepackage[font=normal,labelfont=bf]{caption}

\usepackage{commath}
\usepackage{titlesec}

\journal{Digital Signal Processing}

\graphicspath{{../figure/}}
\begin{document}

\begin{frontmatter}

\title{A CNN based Multifaceted Signal Processing Framework for Heart Rate Proctoring Using Millimeter Wave Radar Ballistocardiography}


\author[add1,add2]{Rafid Umayer Murshed \corref{cor1}}
\ead{rafidumayer.murshed@utdallas.edu}
\cortext[cor1]{Corresponding author}
\author[add1]{Md.Abrar Istiak}
\author[add1]{Md. Toufiqur Rahman}
\author[add1]{Zulqarnain Bin Ashraf}
\author[add1]{Md. Saheed Ullah}

\author[add2]{Mohammad Saquib}

\address[add1]{Department of Electrical and Electronic Engineering (EEE), Bangladesh University of Engineering and Technology (BUET), Dhaka - 1205, Bangladesh}

\address[add2]{Department of Electrical and Computer Engineering, The University of Texas at Dallas (UT Dallas), Richardson, Texas, 75080, USA}





\begin{abstract}
The recent pandemic has refocused the medical world's attention on the diagnostic techniques associated with cardiovascular disease. Heart rate provides a real-time snapshot of cardiovascular health. A more precise heart rate reading enables a better understanding of cardiac muscle activity. Although many existing diagnostic techniques are approaching the limits of perfection, there remains potential for further development. In this paper, we propose MIBINET, a novel multifaceted approach for real-time proctoring of heart rate from \textbf{M}illimeter wave (mm-wave) radar ballistocardiography signals via inter-beat-interval (\textbf{IBI}) using a convolutional neural \textbf{NET}work (CNN). The central theme of our approach is to synergize the feature extraction capabilities of CNN with novel signal processing techniques, resulting in enhanced estimation accuracy while simultaneously reducing computational complexity. This proposed network can be used in hospitals, homes, and passenger vehicles due to its lightweight and contactless properties. It employs classical signal processing prior to fitting the data into the network. Although MIBINET is primarily designed to work on mm-wave signals, it is found equally effective on signals of various modalities such as PCG, ECG, and PPG.  Extensive experimental results and a thorough comparison with the current state-of-the-art on mm-wave signals demonstrate the viability and versatility of the proposed methodology.
\end{abstract}


\end{frontmatter}




\section{Introduction}
\label{sec:introduction}
Human vital signs like Heart Rate (HR), Heart Rate Variability (HRV), Respiration Rate (RR), and Oxygen Saturation (SpO\textsubscript2) are important physiological indicators that reflect the physical and mental well-being of the human body. The heart pumps oxygenated and nutrient-rich blood all over the body. As the cardiac output is intimately associated with HR and stroke volume, HR is central to the cardiovascular process. HR measurement is crucial for a health monitoring system \cite{imp_heart_rate}. According to WHO Global Health Estimates, heart diseases such as myocardial infarction (MI), sudden cardiac death, heart attack, coronary artery disease, arrhythmia, heart valve disease, and heart infection have been the leading causes of death over the last few decades. In 2020, about 697,000 people died alone from heart disease in the United States; one in every five deaths  \cite{centers2019multiple,tsao2022heart}. Clinical biofeedback practice heavily emphasizes the control of cardiac dynamics. To analyze biofeedback, HR, the number of heartbeats per minute is the most often measured metric. However, a healthy heart does not beat uniformly; it changes its rhythm with each beat. While HR focuses on the average number of beats per minute, HRV measures the specific variations in time between successive heartbeats. The period between heartbeats is measured in milliseconds (ms) and is referred to as the \textit{R-R interval} or the \textit{inter-beat interval (IBI)}. The sympathetic and parasympathetic branches of the autonomic nervous system (ANS) control HRV. Especially for diabetic and post-infarction patients, it is a crucial parameter in analyzing the behaviour of the sympathetic and parasympathetic functions of the ANS \cite{Volodina2022-ym}. An effective way of estimating HRV is 
measuring the variation in IBI \cite{rajendra2006heart}, which can be used to detect probable cardiovascular disorders \cite{schuster2016decreased, shaffer2017overview}.

Electrocardiogram (ECG) \cite{malik1996heart, la1998baroreflex} and Photoplethysmography (PPG) are among the recognized methods for measuring HRV indices \cite{lan2018toward}. These systems require the patients to be in constant communication with humans. However, this can be challenging for remote patient monitoring, especially for elderly people living alone or recently discharged patients. Another disadvantage of the aforementioned systems is the limitation of mobility and freedom of the patients. In addition, the electrodes used by ECG and PPG may not only cause unnecessary discomfort but also potentially exacerbate symptoms. Moreover, the palpation of the probes could lead to false alarms and cause alarm fatigue \cite{harrigan2012electrocardiographic,bond2012effects}. Radar-based contactless tele-measurements offer a very comfortable means of continuous vital signal monitoring, which is essential for detecting early onsets of cardiovascular diseases. In this paper, we pursue that line of inquiry.

\par Millimeter wave (mm-wave) based technologies are ideal alternatives for contactless, continuous measurement of human vital signals \cite{yang2016monitoring}. The shorter wavelengths of mm-waves (30–300 GHz) allow for greater spatial resolution. To this end, a number of approaches have been explored to implement mm-wave systems in different biological and medical applications. The most notable applications include contactless measurement of arterial pulses \cite{johnson2019wearable}, cancer diagnosis \cite{topfer2015millimeter, di2017feasibility} and dental diagnosis \cite{hoshi1998application}. In addition, the use of high-resolution mm-wave array beamformers has increased in medical imaging, gesture recognition, and navigation in recent years.
Non-thermal, low-intensity electromagnetic radiation is used in mm-wave therapy (MWT), a novel and revolutionary method \cite{6338252}  of treating patients. The implementation of mm-wave FMCW radar \cite{8695699} is used to measure breathing and HRs. To enhance the effectiveness and feasibility of these technologies in real-time, their implementation must take into account both computational efficiency and optimal precision.
The above studies do not analyze both the computational efficiency and optimal precision of the used mm-wave technologies simultaneously. Unlike those, our proposed mm-wave-based technique, MIBINET, can estimate the IBI values with optimal precision in real time.

\begin{figure}
\centerline{\includegraphics[width = 0.5\textwidth]{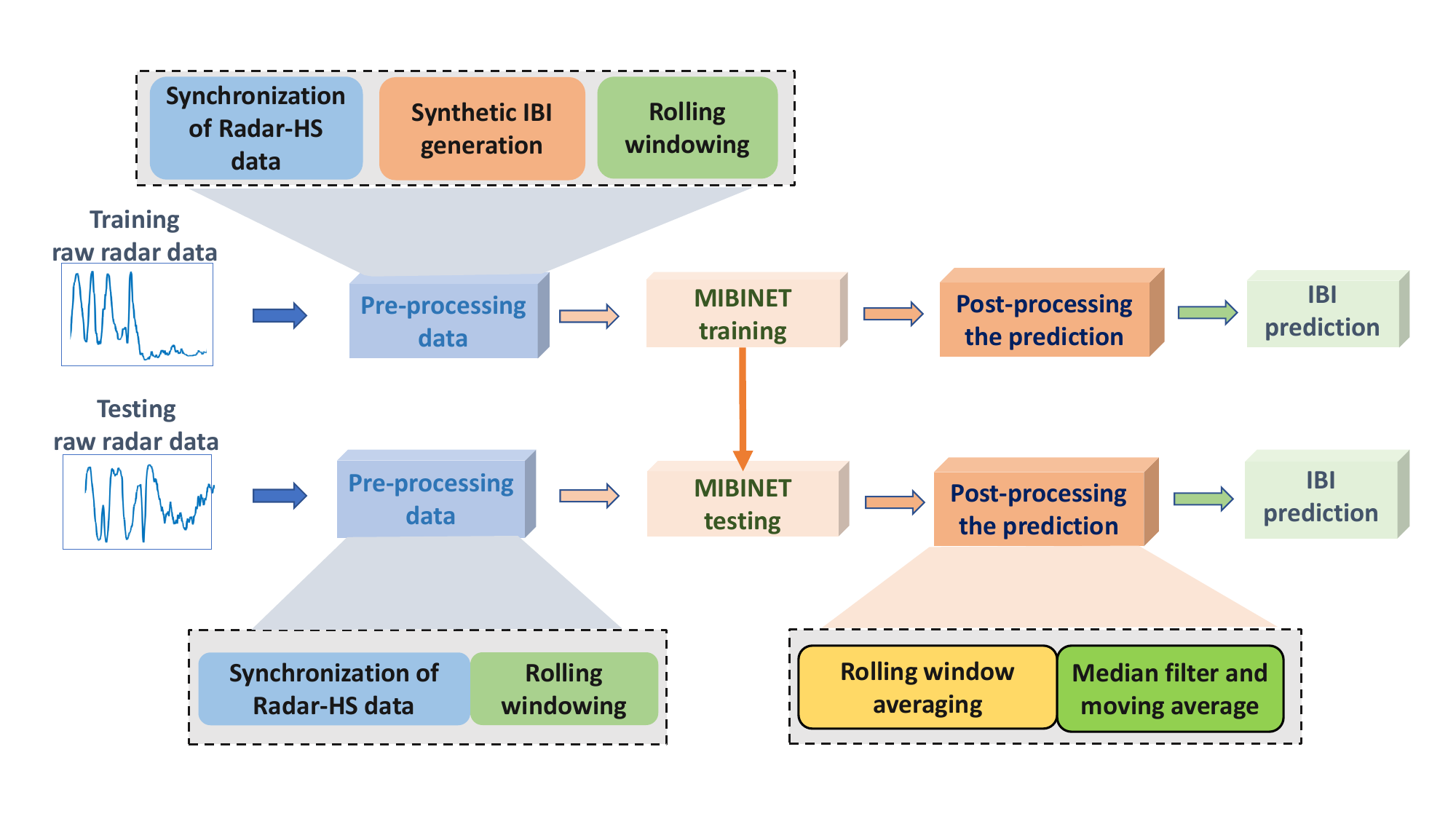}}
\caption{Overall workflow of the proposed approach.}
\label{overall_scheme}
\end{figure}

To the best of our knowledge, MIBINET is the first neural network-based approach that operates on pre-processed mm-wave data to estimate the instantaneous IBI. Our work's fundamental premise is to improve estimation accuracy and computational complexity by fusing the feature extraction capabilities of CNN with distinctive signal processing strategies. It adopts a rolling window-based pre-processing approach to confine the input data to a predetermined size for conveniently feeding into a neural network's input layer. The network's output is also reprocessed with a novel rolling window averaging approach followed by various traditional post-processing filters to improve the estimation accuracy; see Figure \ref{overall_scheme}. Our numerical results demonstrate that the proposed MIBINET is capable of outperforming the state-of-the-art techniques in terms of IBI estimation precision by more than 5\%. Below is a summary of the other major contributions of this investigation:

\begin{enumerate}


\item We introduce a synthetic IBI augmentation technique to enrich the dataset, significantly enhancing the correlation coefficient and reducing the root-mean-square error (RMSE). This data augmentation enables our method to demonstrate robust performance across various patients and even in anomalous cases. 
\item Our approach combines a rolling window-averaging technique with various traditional post-processing filters to improve the estimation accuracy, resulting in a more than 5\% increase in IBI estimation precision compared to state-of-the-art techniques.
\item We employ a custom loss function specifically designed to reduce outliers, which leads to substantial improvements in model performance. Our designed lightweight 1D-CNN model architecture facilitates real-time use and exhibits high performance across 11 different test subjects.

\end{enumerate}

The rest of this paper is organized as follows: Section \ref{rel_work} contains a review of the current state of contact-free vital sign monitoring. MIBINET is proposed in Section \ref{methodolog}, along with the description of pre and post-processing techniques. Section \ref{result} includes the numerical results. Finally, a brief discussion of the findings, limitations, and strengths of MIBINET are provided in Section 5. Regarding notation, scalars and vectors are represented in lower and bold lower cases, respectively. The $n^{\mathrm{th}}$ element of vector $\mathbf{a}$ is denoted by $\mathbf{a}(n)$. 

\section{Related Works}
\label{rel_work}
A significant amount of research has focused on non-invasive contactless monitoring of vital signs (e.g., HR, RR, HRV) due to its many advantages. Non-invasive and contactless optical approaches for heartbeat monitoring based on optical Doppler interferometry and laser have been proposed in \cite{J_Bancifra2022-xy,4286932}. Here, the photo-EMF pulsed laser vibrometer (PPLV) is studied, where the subjects were instructed to first exhale and then hold their breath for as long as they could after inhaling. This is to avoid any potential muffling of the cardiac signals by the subjects' respirations. However, it might cause unnecessary discomfort to the patients. Fengyu Wang et al. \cite{wang2021mmhrv} proposed contactless HRV monitoring using mm-wave radio. First, they developed a user-locating target detector without calibration. Heartbeat signal extractors then optimize the decomposition of chest-movement-modulated channel information to uncover the desired signal. Now the pulse signal's peak position can be used to evaluate HRV parameters for each target utilizing the IBI values. This method, mmHRV, can assess HRV with a median IBI error of 28 ms (w.r.t 96.16\% accuracy) for 11 players in the line of sight (LOS). For non-LOS, it is 31.71 ms. However, this proposed method exhibited considerably uneven errors among the participant subjects. Zhang et al. \cite{zhang2020health} suggested radio signal-based contactless MI detection. This work establishes MI detection using RF signals, providing contactless, non-intrusive, continuous home monitoring for MI hazards. They also proposed heartbeat signal segmentation and MI detection algorithms. Extensive evaluations have been conducted to confirm the effectiveness of Health-Radio. However, this approach has not been generalized to work over different signal modalities such as PPG, ECG, and PPG.

\begin{figure}[b]
\centerline{\includegraphics[width = 0.46\textwidth]{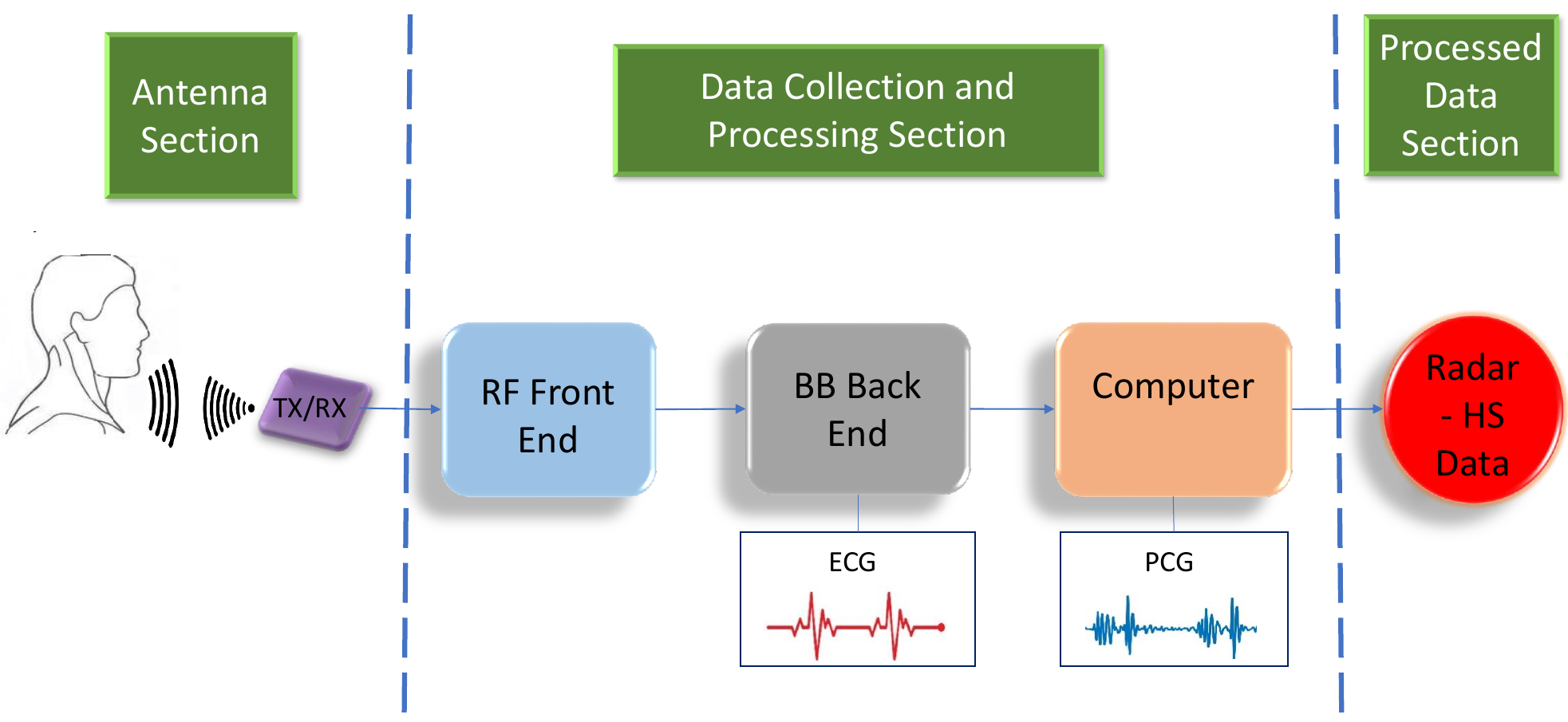}}
\caption{Data collection scheme - The figure depicts how the data was collected.}
\label{fig_1}
\end{figure}

Radar technology is one of the most promising possibilities for contactless, non-invasive monitoring of vital signs such as cardiac signals. Impulse radio ultra-wideband (IR UWB) radar was used to monitor vital signs \cite{8378778,8310011}. While monitoring, this IR UWB-based technique exploits signal properties, completely disregarding any object qualities with which this signal interacts. Body-coupled antennas and UWB pulsed radar in-body monitoring of lungs and heart motion are used in this scenario. The radar has to be installed at a specific location. Moreover, sudden oscillations cause abrupt phase fluctuations, which impact HR predictions. Many methods for calculating HR \cite{8906072,8887436} from UWB radar data have recently been developed. Their performance, however, is still insufficient for practical applications. Frequency-modulated continuous wave (FMCW) radar \cite{8378778}, continuous wave (CW) Doppler radar for HR monitoring, and respiration monitoring have been developed significantly during the past few years \cite{article,5247107,8732355}. Because CW radar or Doppler radar does not capture the target's range or distance, FMCW radar is designed to overcome this issue due to its range and radial velocity measurement. Since FMCW radar operates at lower transmit power, the received signal will not only be distorted by the environment but also could be weak. Although beamforming and range-gating approaches can separate the signal of interest from the noise, several difficulties, such as random body motions, must be addressed before radar-based non-contact measurements can be implemented in real-world applications. Sakamoto, Takuya, et al. \cite{sakamoto2015feature} addressed the use of a UWB radar system for estimating the human HR precisely. The performance of the proposed approach was demonstrated through measurements. Target classification and type recognition are achievable with UWB radar because the received signal contains information not just regarding the target as a whole but also about its individual component elements \cite{1006348}. The suggested approach estimates the HR efficiently and correctly by using the feature points of a radar signal. Nonetheless, as heartbeat waveforms vary, even within the beat-to-beat interval, Fourier and periodicity-based approaches are ineffective for estimating instantaneous HR in real-time. Consequently, a Wavelet-Transform (WT) based method \cite{7997608} has been proposed for faster HR detection using Doppler radar because of the insufficient frequency resolution of the Fourier transform (FT). Unfortunately, the WT approach requires extensive signal processing in order to identify HR.
\par In 2010, researchers started applying machine learning (ML) in the field of ballistocardiography signals. Bruser et al. \cite{bruser2010applying} brought an idea of an unsupervised modified K-means clustering training algorithm to estimate parameters for the BCG signal. Then, the parameters followed by the heartbeat detection and refinement process combine three indicators for localizing the heartbeat. Collectively, the method called BEAT has to be re-trained when facing a significant change in the BCG signal. Later, a real-time approach for identifying individual heartbeats without requiring intensive signal processing was studied in \cite{s20082351}. It is an Artificial Neural Network (ANN) based Heartbeat Detection Technique using a state-of-the-art mm-wave radar sensor. The ANN is trained on the raw radar signal, offering computational simplicity while estimating IPIs (inter-pulse intervals) with relatively low accuracy. Hence, a signal of high HRV is not suitable for this study. In addition, this shallow ANN model has a 2\% chance of missing a pulse. Nowadays, adopting deep learning methods enables the proliferation of automatic, portable, non-invasive data-driven HR monitoring. In 2019, Dwaipayan et al. \cite{8607019} proposed a temporal model CorNET (CNN + LSTM) method to predict HR and biometric identification. This was evaluated on two subjects and consequently lacked subject bias. This model needs longer training in order to adapt to the HR variability in an ambulant environment. \footnote{Unlike \cite{8607019}, we tackled variation with augmentation and rolling window averaging techniques.} Xiangmao et al. \cite{chang2021deepheart} proposed a method to estimate HR after acquiring clean input PPG data from a denoising CNN (DCNN). Although DCNN adds more robustness to real-life artefacts, it adds overhead in time complexity. On a different note, the first triumphant attempt to predict pulse rate (PR) from facial video data after the surge of computer vision was the proposition of PRnet \cite{huang2021novel}. This method leverages the synergy of 3D convolution and LSTM to predict pulse rate from spatiotemporal features with less error. It requires a minimum of 2 seconds (60 frames) to effectively measure PR, which limits the method from being a real-time pragmatic solution.

\section{Methodology}
\label{methodolog}
In this section, we describe the working principle of MIBINET. Since it is a DNN based-technique, it needs to be trained and validated on labelled data. First, we describe how the labelled data is collected.

\subsection{Data Acquisition}



Figure \ref{fig_1} is a representation of the data collecting and processing procedure following the work presented in \cite{shi2020dataset}.
Shi et al. \cite{shi2020dataset} made the dataset available in 2020 to foster the field of radar heart sound (Radar-HS) in a contactless manner. The dataset was approved by the ethics committee of the Friedrich-Alexander-Universität
Erlangen-Nürnberg for maintaining guidelines and regulations. Their tailored hardware setup initiates from the RF front end as shown in Figure \ref{fig_1}, consisting of a six-port, and it is utilized as a quadrature interferometer for radar applications. The six-port has two input signals and four output signals, where two input signals consist of a reference signal at a defined frequency and a received signal reflected from the target. The antenna direction was set perpendicular to the test subject's thorax surface to maximize the signal quality. The baseband board back end (BB Back End) receives radar signals from the RF front end and digitizes the signals. It is equipped with an ECG and respiration sensor (RS) for simultaneous sampling. The digitized raw signals were subsequently received by a computer equipped with a PCG that served as a reference sensor. PCG signals are stored after re-sampling and synchronization.

\begin{figure}[h]
    \includegraphics[trim={65 425 65 72},clip,width=\linewidth]{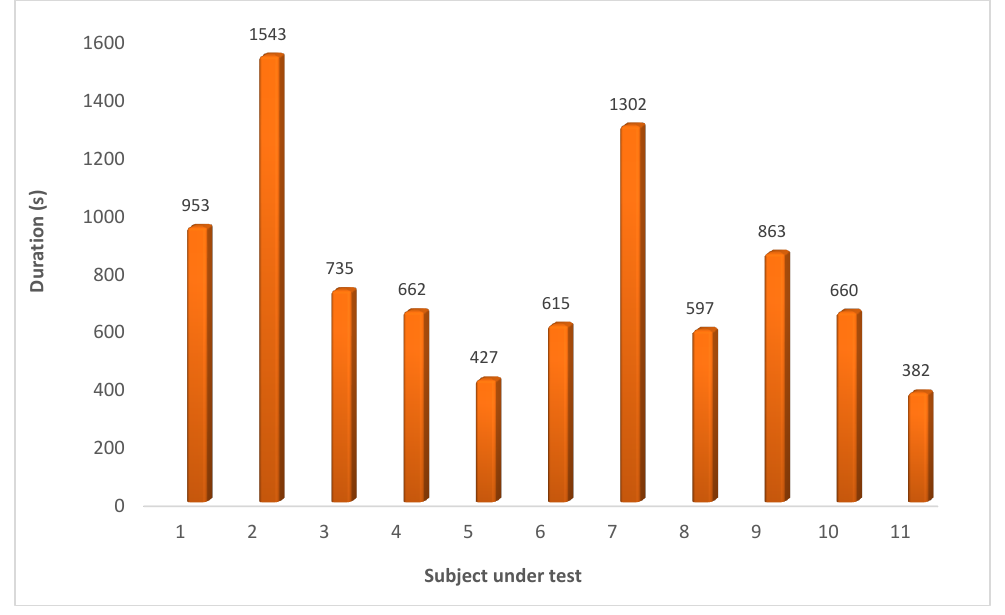}
    \caption{Total signal duration of each subject in the dataset.}
    \label{duration_bar}
\end{figure}

Let us now take a brief glance at the processed data, which consists of the raw radar data along with the filtered data from the respiration sensor, the ECG leads, the PCG data and the Radar heart sounds. Figure \ref{duration_bar} provides a breakdown of the collected data. As we can see, 11 subjects participated in this measurement process. Although the total data consist of 13376 s of recordings, only the default scenarios are of interest to us as Shi et al. \cite{shi2020dataset} made them publicly available. In the following sub-section, we describe our adopted rolling window-based pre-processing scheme, which allows us to seamlessly meta-morph the collected dataset in a form compatible with deep neural networks.

\subsection{Dataset Distribution}
\label{Dat_dist}

Following standard practice, the entire dataset was permuted into 11 folds, with each subject's data appearing only once in the test set of each fold. Each fold consists of separate training, validation, and test sets. The folds are numbered according to the ID of the subject in the test set. For instance, fold one indicates that the data of the first subject is in the test set. The training and validation sets consist of the data of the remaining ten users, with eight in training and two in validation.
\begin{figure*}[t]
\centerline{\includegraphics[width = \textwidth,trim={0cm 3.5cm 0cm 3cm}]{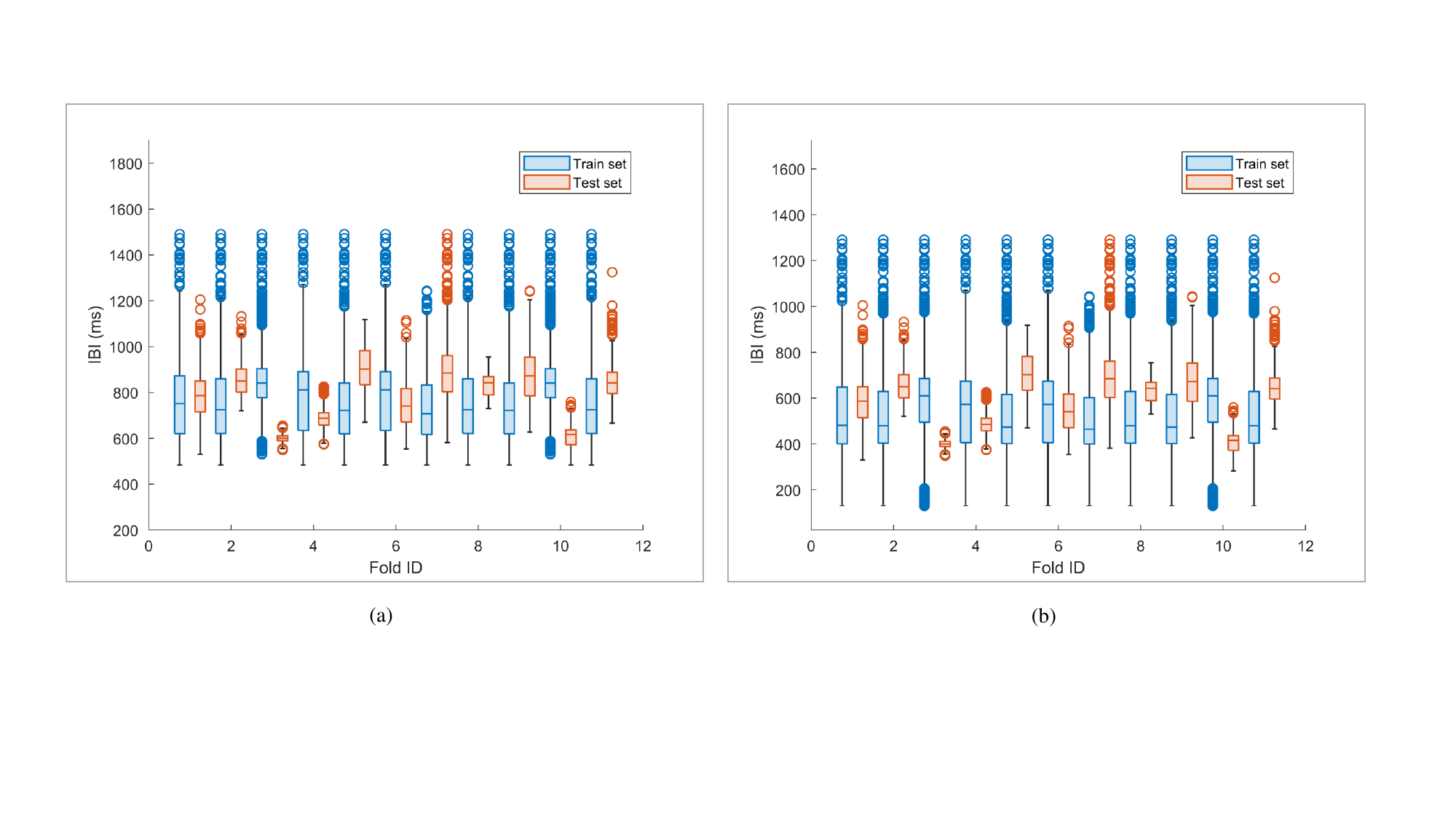}}
\vspace{-6mm}
\caption{Distribution of IBI values in box plot: (a) before augmentation and (b) after augmentation.}
\label{IBI_dist}
\end{figure*}

The distribution of the IBI values from the data of different test subjects reveals the typical variations usually found in clinical trials. For example, as shown in Figure \ref{IBI_dist}a, the median values of the participant test subjects vary from 0.6 to 0.8 seconds (s), with subjects 3 and 10 having comparatively lower IBI values while subjects 7, 8, and 9 exhibit relatively high values.

Analyzing the distribution of each fold shows some disparities between the range of IBI values in the training and test sets. In particular, folds 3 and 10 display significant variations, making it quite challenging to train deep learning-based models. This leads us to propose a novel data augmentation approach to training the deep models, which will be elaborated on next.

\subsection{Sythetic IBI Generation Augmentation}
 
\begin{figure*}
\subfloat[]{
 \includegraphics[height=75mm,width=90mm]{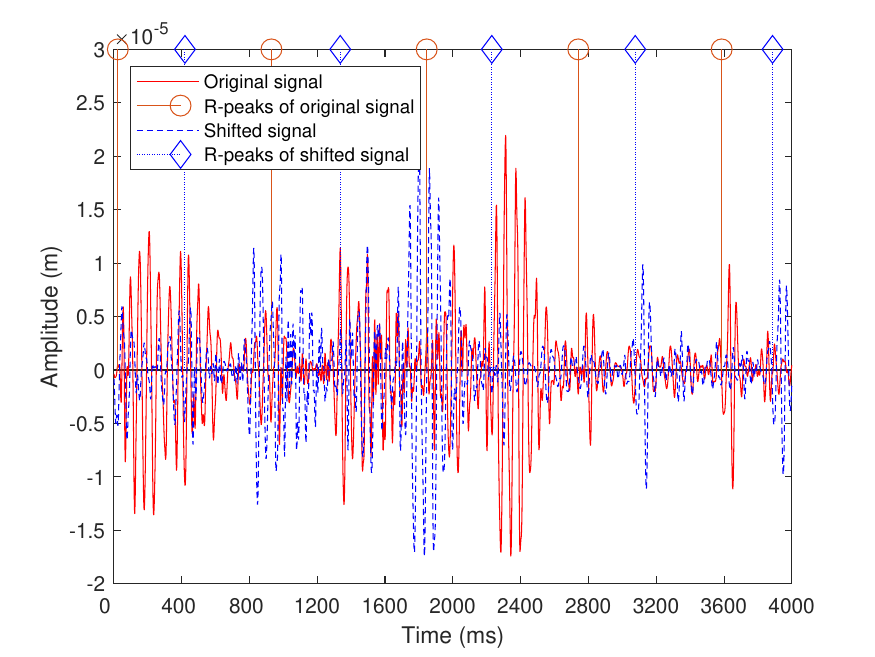}

  \label{Superposition_Aug}
}
\subfloat[]{
\includegraphics[height=75mm,width=90mm]{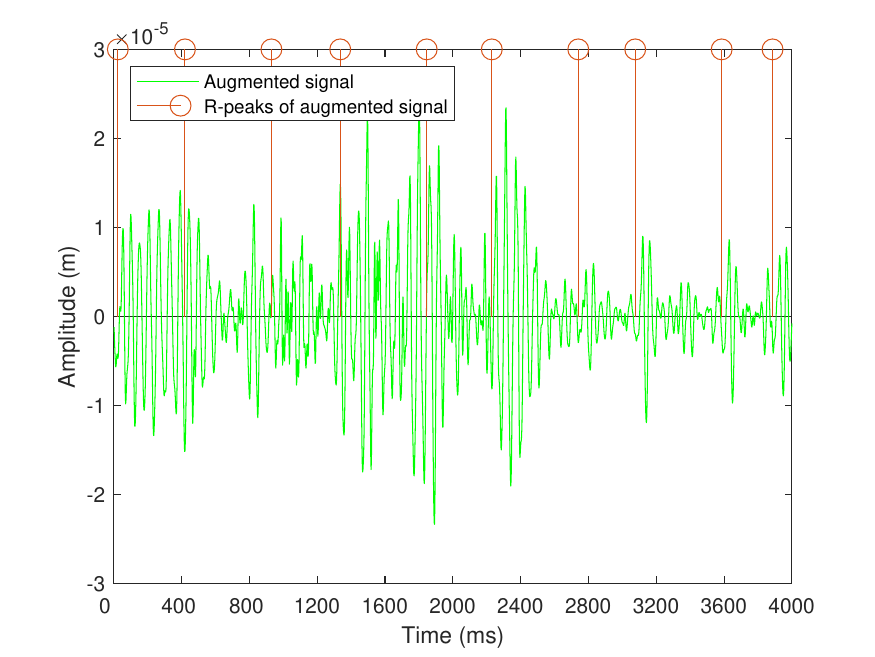}
  \label{Aug_signal}
}

\caption{Synthetic IBI augmentation: (a) superposition of signals for augmentation and (b) augmented signal with new R-peaks.}
\label{time_domain}
\end{figure*}

Due to the scarcity of lower IBI values in the training sets of folds 3 and 10 (from Figure \ref{IBI_dist}a), the model struggles to perform well on the test sets, which contain a relatively lower range of IBI values. We resolve this issue by generating synthetic radar-HS signals to inject smaller IBI values into the training set. To synthesize radar-HS signals characterized by smaller IBI values, specifically in the range of $0.5-0.6$ s, we first start with signals that have IBI values in the higher range, preferably in the region of $0.9-1.1$ s. We then left-shift those signals in the time domain by $0.4-0.5$ s and add them with their shifted versions as depicted in Figure \ref{Superposition_Aug}. Mathematically, it can be expressed as follows

\begin{equation}
\label{augmentation}
  x_{\mathrm{aug}}(t) = 
  { x_{\mathrm{o}}(t) + x_{\mathrm{o}}(t + \alpha)},
\end{equation}
where $x_{\mathrm{aug}}(t)$ and $x_{\mathrm{o}}(t)$ denote the augmented and original signals respectively. Here, $\alpha$ is a uniform random variable between 450 ms and 550 ms. Now the R-peaks of the augmented signal are the union of the R-peaks from the original signal and the left-shifted signal, as shown in Figure \ref{Aug_signal}. Due to such augmentation, the distribution of the resulting training set becomes much more consistent, as can clearly be observed in Figure \ref{IBI_dist}b. In brief, this IBI augmentation technique synthesizes comparatively smaller IBI values corresponding to elevated tachycardia heart rates. Similarly, right-shifting the signal can make the distribution adaptable for high IBI bradycardia conditions. It is important to note that this IBI augmentation technique is particularly effective for periodic signals, ensuring the correct deployment of the method while maintaining the accuracy of the results.

In order to further facilitate our neural network-based approach, we supplement the processed signals with an additional rolling window-based pre-processing scheme, which will be discussed next.

\subsection{Rolling Window-based Pre-processing and Augmentation}

\begin{figure*}[t]
    \includegraphics[trim={30 20 30 38},clip,width=\linewidth]{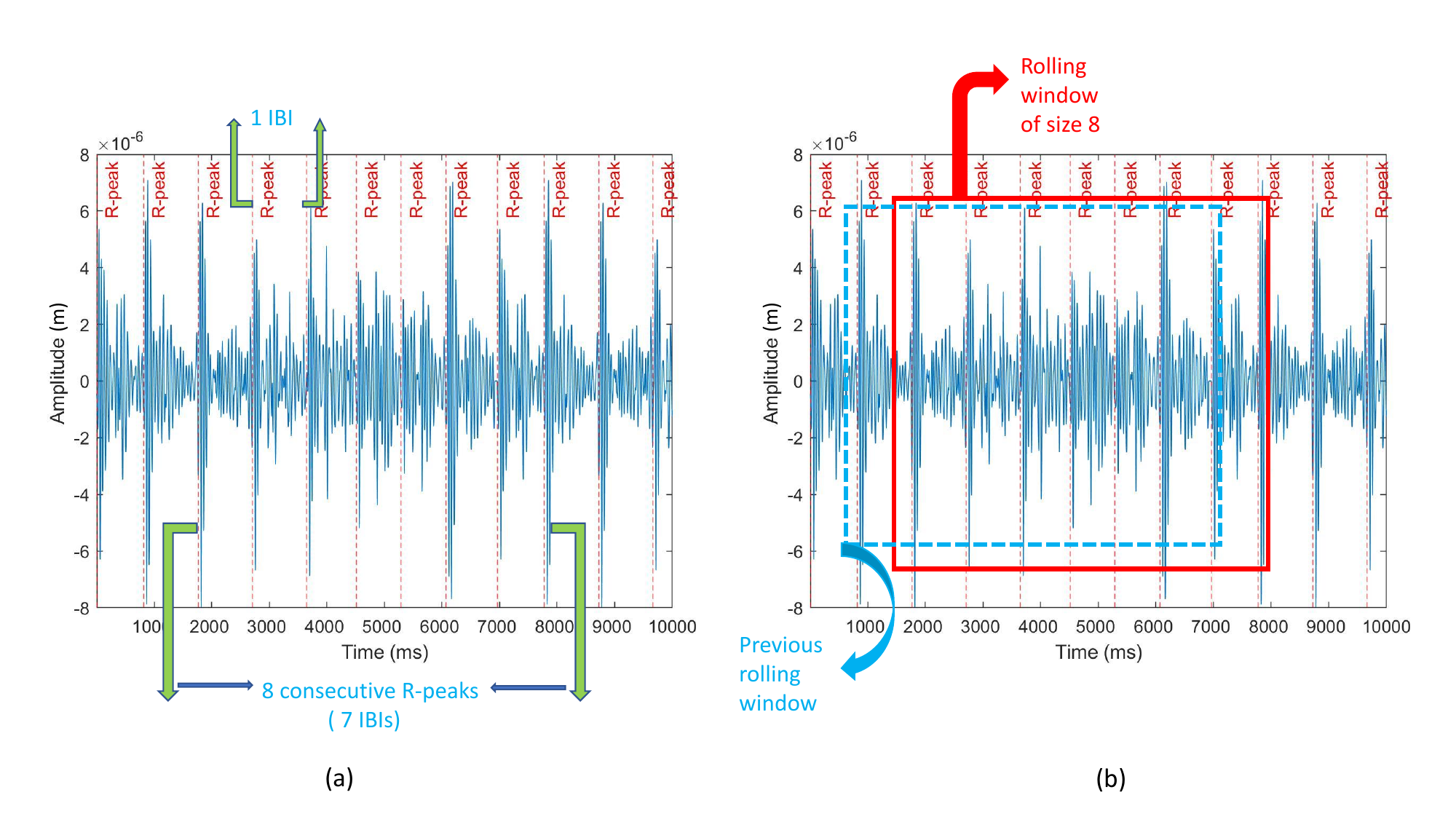}
    \caption{Rolling window-based pre-processing: (a) original data and (b) windowing of data.}
    \label{rolling_pp}
\end{figure*}

IBI is the time interval between two consecutive R-peaks, as shown in Figure \ref{rolling_pp}a. It can be noticed that the rolling window extends over 8 consecutive R-peaks of the processed signals, each containing 7 IBIs; see Figure \ref{rolling_pp}b. Although the specific size of the rolling window is not based on exact theoretical calculation, our empirical results demonstrate that the proposed size is near optimal for this study to trade-off between computational time and performance. 
Typically, most neural networks accept predefined dimensions of the inputs and outputs. However, the available radar data does not adhere to those constraints. Hence, we restrain the dimensions of the inputs and outputs to a predefined value. Getting inspired from \cite{jana20201d, panwar2020pp}, by using a rolling window of size 8, we are able to restrict the size of the output to a vector of size 7. In our dataset, each input consists of a windowed version of the original data containing 7 IBIs, and the maximum windowed input length is found to be 4885 samples. In order to restrict the dimension of the input data, we zero-pad all the inputs to a size of 4910 samples. These zeroes are inserted randomly at the start and end of each window to remove any prospective bias during training. The starting point of every successive rolling window is randomly selected between two consecutive R-peaks to remove any regional bias.

\subsection{Architecture of the MIBINET}

Our proposed CNN architecture aims to balance computational efficiency and high prediction accuracy for estimating IBI values. The architecture, named MIBINET, incorporates fewer parameters than other contemporary networks while maintaining high accuracy. The rationale behind developing such an architecture is to create a lightweight and efficient real-time heart rate monitoring solution.

MIBINET combines standard spatial convolutional layers and depth-wise separable convolutional layers to extract distinguishable features from the one-dimensional input data. The intuition behind using convolutional layers first is to exploit their ability to capture local patterns and spatial dependencies in the input data, enabling the extraction of meaningful features. To ensure faster convergence during training, the input data is first normalized and passed through an initial batch normalization layer that adjusts the mean and variance of the input. Following this preprocessing step, the network comprises a sequence of convolutional and pooling layers with varying kernel numbers and sizes. We have determined these kernel parameters through extensive experimentation and trial-and-error to optimize feature extraction. The reasoning for employing pooling layers following the convolutional layers is to diminish the spatial dimensions of the feature maps while preserving crucial information, thereby enhancing computational efficiency and mitigating overfitting by reducing the number of parameters.

\begin{figure*}[t]
    \includegraphics[trim={70 25 70 15},clip,width=\linewidth,height = 9.6 cm]{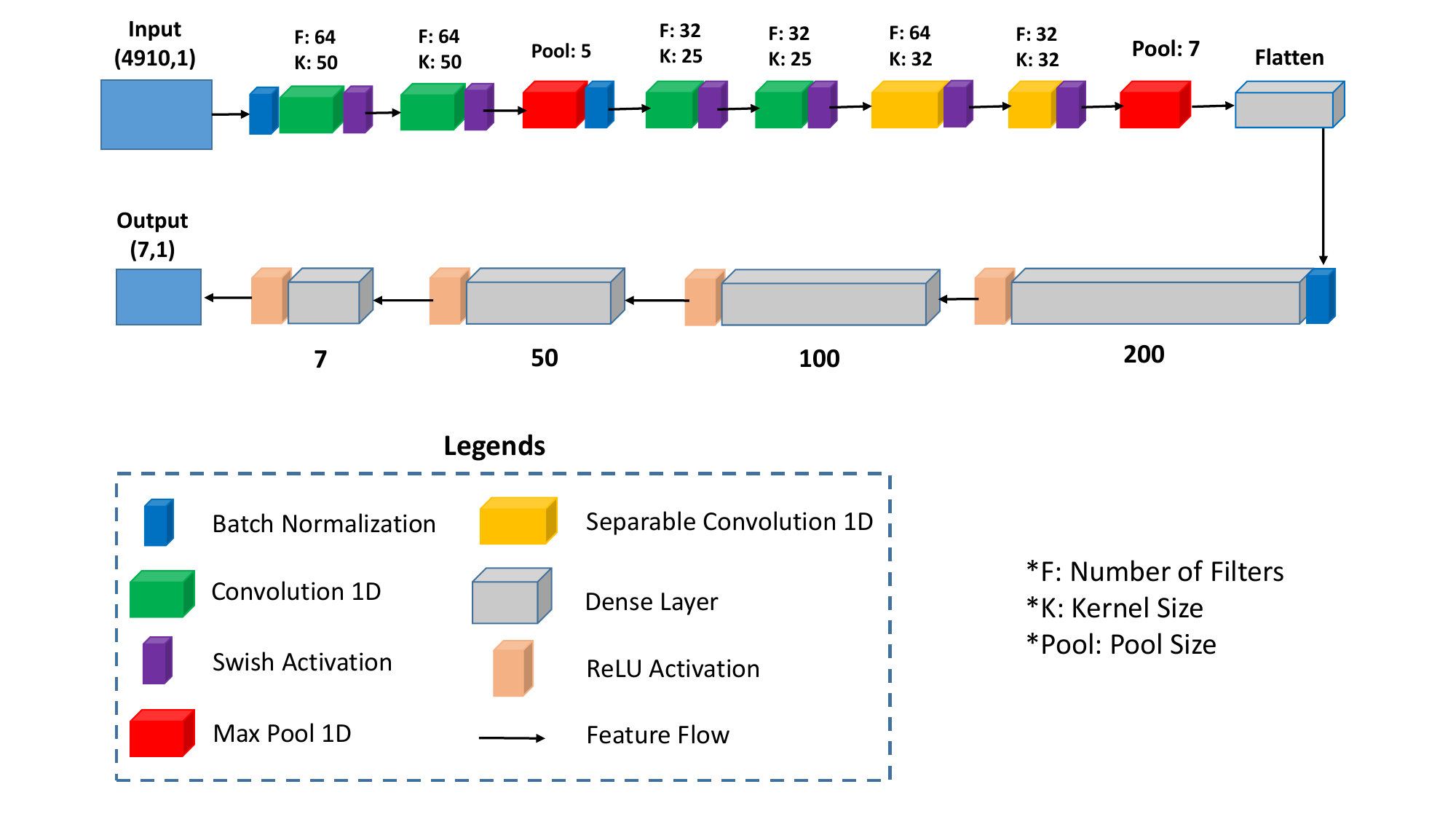}
    \caption{Proposed MIBINETs' 1D-CNN architecture.}
    \label{architec}
\end{figure*}

Subsequently, the extracted features are fed into a fully connected network of dense layers. The use of fully connected layers after the convolutional layers allow for the combination and processing of the extracted features at a higher level, ultimately leading to the final prediction. The final dense layer predicts an array of length seven, representing consecutive IBI values. We employ the swish activation function \cite{swish} in the convolutional layers, as it outperforms ReLU in this context, while ReLU is used in the fully connected dense layers. To further enhance computational efficiency, MIBINET utilizes multiple 1D depth-wise separable convolutional layers, reducing the number of parameters involved. A schematic diagram of the entire model architecture is illustrated in Figure \ref{architec}.

\subsection{Custom Loss Function}
\label{loss_section}
The loss function plays a vital role in supervised learning algorithms such as feed-forward networks. It determines how the predictions approach the target labels. Since our goal is to measure the IBI values consistently with high precision, we not only require the error between the predictions and true values to be small but also want there to be a high correlation between the predicted values and the true values. To achieve this, we design a novel weighted loss function that takes into account the aforementioned considerations. The designed loss function is a weighted sum of three well-known loss functions for regression tasks, namely, the mean squared error (MSE) loss, the Huber loss (HL) \cite{huber1992robust}, and the mean absolute error (MAE) loss. While the MSE is too sensitive to outliers, the MAE weights all the errors equally, disregarding outliers completely. The HL provides a good balance between both the MSE and the MAE. 
Finally, to ensure a high correlation between the predicted outputs and the actual value, we add another component to our weighted loss in the form of correlation coefficient loss. The resultant loss function can be expressed as

\begin{equation}
\label{prop_loss}
{\mathcal{L}_{\mathrm{corr}} = w_1\times(1-r^{2}) + w_2\times L_{\mathrm{HL}} + w_3\times \epsilon^{2} + w_4\times |\epsilon| } + w_5\times L^a_{\mathrm{HL}}
\end{equation}
where $L_{\mathrm{HL}}$ denotes the well-known Huber-loss value, $L_{\mathrm{HL}}$ is a modified asymmetric version of the Huber-loss, $r$ represents the Pearson correlation coefficient, $\epsilon$ is the error in the prediction, and $w_1,w_2,w_3$, $w_4$ and $w_5$ are the weights applied to the correlation loss, Huber-loss, MSE, MAE, and the asymmetric HL, respectively. The exact values used for $w_1,w_2,w_3$, $w_4$ and $w_5$ are determined through extensive trial and error to best suit the required task. It is also noteworthy that none of the well-known regression losses (i.e., MSE, MAE, mean absolute percentage error, etc.) except HL solely performed well in our experiments. Here, for each value $\epsilon$ in error = $y$\textsubscript{pred}-$y$\textsubscript{true}, $L^a_{\mathrm{HL}}$ is given by 

\begin{equation}
L^a_{\mathrm{HL}} = \begin{cases}
          0.5\times (\epsilon^2 + |\epsilon| + \rho \times \epsilon) \quad &\text{if} \, |\epsilon| \le \psi \\
          2\times \psi \times |\epsilon| + \rho \times \psi \times \epsilon - 0.5 \times(\psi)^2 \quad &\text{if} \, |\epsilon| > \psi \\
     \end{cases},
\end{equation}
where `$\psi$' is a constant parameter that determines the exact behaviour of the loss towards outliers, and $\rho$ is either -1 or +1, depending on whether we want positive or negative asymmetry. We choose a typical value of $\psi$ = 1 for all simulations. The value of $\rho$ is chosen to be +1 since the dataset had more high IBI values.
\subsection{Post-Processing}
\label{pp}
We first process the outputs by a rolling window averaging scheme to smooth out the predictions. This exploits our rolling window-based pre-processing scheme, where there is considerable overlapping between neighbouring window samples. More precisely, each window retains seven R-peaks from the previous window and hence has exactly six same IBIs. This is utilized during the post-processing, where all predictions on the same initial IBI value are averaged, and hence the errors are further reduced. Following the post-processing schemes of \cite{shi2020dataset}, we also use a median filter of length 5 and a moving average filter of length 6, which boosts the reliability of the model’s predictions.

\begin{figure}
  \includegraphics[trim={80 40 40 10},clip, width = \linewidth]{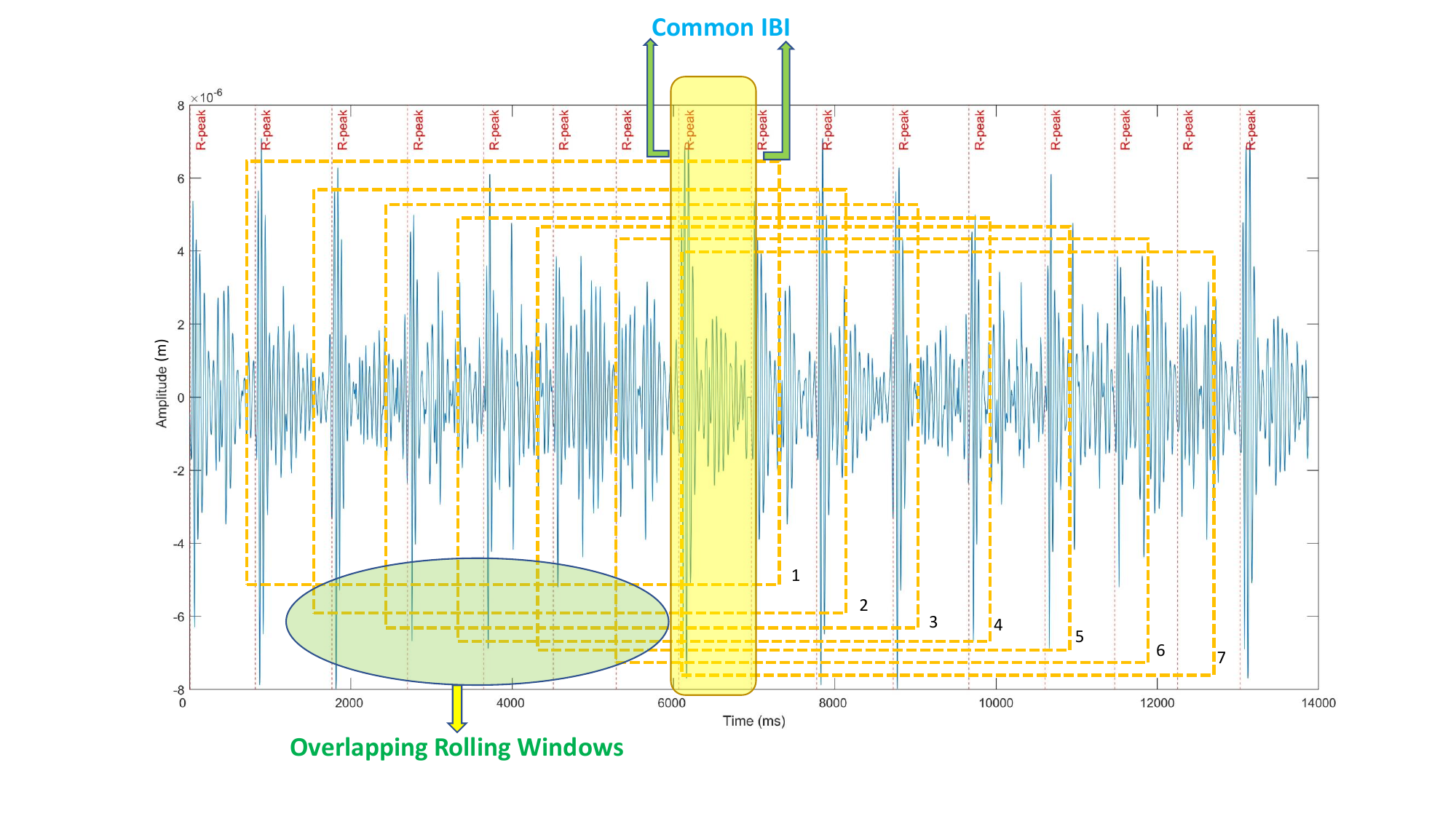}
  \caption{Rolling window averaging: Illustration of seven overlapping successive windows and their intersection.}
  \label{rolling_avg}
\end{figure}

As described earlier, each rolling window consists of seven consecutive IBIs; see Figure \ref{rolling_pp}. Consequently, this leads to the overlapping of successive windows, and the intersection of each of seven such successive windows contains a single common IBI as shown in Figure \ref{rolling_avg}. Furthermore, our model predicts the IBI values from each window separately. Thus, it gives seven distinct predictions for each IBI value. Averaging those IBI values yields a more accurate prediction. Mathematically, this can be expressed as

\begin{equation}
\label{correlation_coeff_loss}
  z(k) = \frac{ \sum_{i=1}^{7}\mathbf{y}_{k-7+i}(8-i)}{
        7};\quad k = 7,8,\ldots
\end{equation}
where $z(k)$ denotes the $k^{\mathrm{th}}$ IBI value of a sample and $\mathbf{y}_j$ contains seven consecutive IBI predictions from the $j^{\mathrm{th}}$ rolling window. 


\section{Experimental Results}
\label{result}
In this section, we discuss the experimental data and analysis used to evaluate the performance of our proposed MIBINET. We assess the efficacy of our proposed system using the Pearson correlation coefficient and root means square error (RMSE). These metrics are chosen due to their suitability for regression tasks. In addition, the established state-of-the-art in \cite{shi2020dataset} is used as the baseline methodology.









\subsection{Evaluation Metrics}
The expression of the Pearson correlation coefficient is
\begin{equation}
\label{pearson_correlation_coeff}
  r = 
{\frac{ \sum_{i=1}^{n}\left(x_i-\bar{x}\right)\left(y_i-\bar{y}\right)}{ \sqrt{\sum_{i=1}^{n}\left(x_i-\bar{x}\right)^2}\sqrt{\sum_{i=1}^{n}\left(y_i-\bar{y}\right)^2}}}\
\end{equation}
where $x_i$ and $y_i$ are the $i^{\mathrm{th}}$ predicted IBI and the ground truth IBI, respectively. Their sample means are denoted by $\bar{x}$ and $\bar{y}$. The term $n$ denotes the total number of IBIs in the sample. The other metric, RMSE, can be defined as
\begin{equation}
\label{rmse_eqn}
  \mathrm{RMSE} = 
  \sqrt{\frac{ \sum_{i=1}^{n}(x_i-y_i)^2}{n}}\,.
\end{equation}

It is to be mentioned that all 11 users' ground truth and predicted IBI are concatenated prior to the evaluation of the final results.

\subsection{Training Setup}

\begin{figure}[t]
    \includegraphics[width=\linewidth]{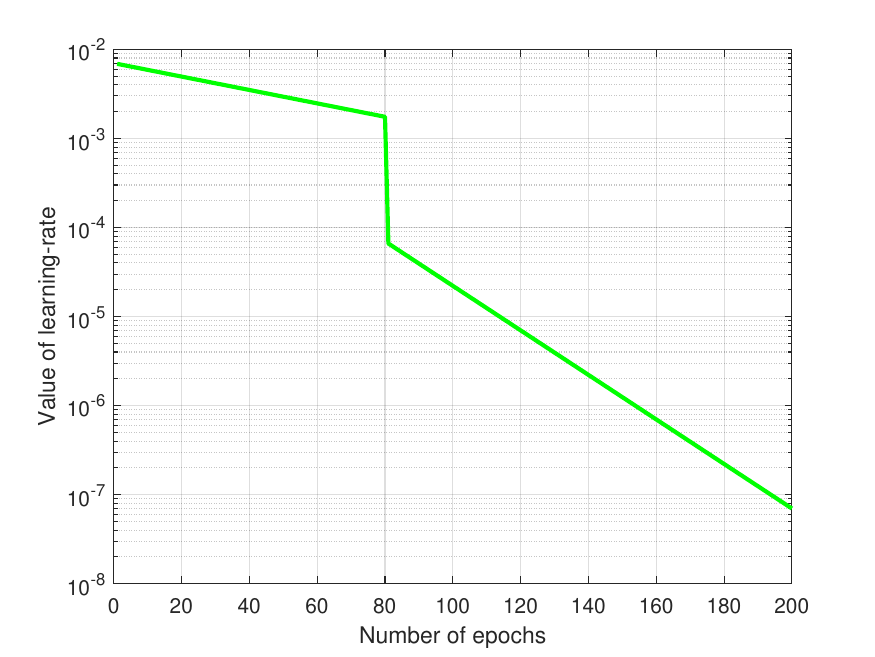}
    \caption{Learning rate used for MIBINET training with the number of epochs.}
    \vspace{-2mm}
    \label{learningrate}
\end{figure}

All simulations are conducted using Google Colaboratory, a Python development environment that runs in the browser using Google Cloud and provides free access to powerful graphical processing units (GPU). Our proposed MIBINET and peripherals are implemented in Python 3.7 utilizing TensorFlow 2.9 and a Tesla T4 GPU provided by Google Collaboratory. All codes are executed with the same setup to enable an accurate and fair comparison of different processes. We have trained our network by utilizing a batch size of 1024 for 200 epochs. The learning rate is tuned using an exponential decay scheduler as shown in Figure \ref{learningrate}. The initial learning rate is set to 0.007. It is decreased by half every 40 epochs to 0.0018 after 80 epochs. Then, it is set to 0.00007 and is decreased by 10 times every 40 epochs through the next 120 epochs. All experiments used Adam \cite{kingma2014adam} optimizer along with momentum with a decay of 0.9. In addition, the best weights were saved based on the validation-weighted metric. The weighted metric can be expressed as
\begin{equation}
\label{Weighted_metric}
{\mathcal{M}_{\mathrm{weighted}} = \alpha_1\times(1-r^{2}) + \alpha_2\times \epsilon^{2} + \alpha_3\times  |\epsilon| },
\end{equation}
where the weights $\alpha_1$, $\alpha_2$, and $\alpha_3$ are chosen to be 10, 0.1, and 0.1, respectively, to ensure equal importance on correlation coefficient and RMSE. Final model predictions are compared after the post-processing as mentioned in sub-section \ref{pp}. It is to be noted that, for every fold, the model was trained with the same parameters. In all our simulations, we have set $w_1 = 0.002$, $w_2 = 1.0$, $w_3 = 0.0096$, $w_4 = 0.002$, and $w_5 = 0.0032$ in (\ref{prop_loss}) as those are found to be near optimal.

\subsection{MIBINET vs. HSMM Method}

The Radar-HS dataset in \cite{shi2020dataset} has recordings sampled at multiple frequencies. To maintain consistency, all the recordings are resampled to 500 Hz. We then prepare the dataset for MIBINET as described in Section \ref{Dat_dist}. Following the aforementioned pre-processing and post-processing steps, the final results are generated, and the fold-wise results are given in Table \ref{foldwise_perform}. The fold-wise results of the HSMM method, the existing state-of-the-art, are also provided here for ease of comparison.

\begin{table}[h]
\caption{Fold-wise performance}
\label{foldwise_perform}
\resizebox{\linewidth}{!}{
\begin{tabular}{ccccc}
\hline
\begin{tabular}[c]{@{}c@{}}Test\\ User No\end{tabular} &
  \multicolumn{2}{c}{Correlation Coefficient, $r$ (\%)} &
  \multicolumn{2}{c}{RMSE (ms)} \\ \cline{2-5} 
 &
  HSMM Based \cite{springer2015logistic} &
  \multicolumn{1}{c|}{MIBINET} &
  HSMM Based \cite{springer2015logistic} &
  MIBINET \\ \hline
1 &
  \cellcolor[HTML]{FFFFFF}78.41 &
  \multicolumn{1}{c|}{\textbf{97.08}} &
  \cellcolor[HTML]{FFFFFF}59.58 &
  \cellcolor[HTML]{FFFFFF}  \textbf{19.77} \\
2 &
  \cellcolor[HTML]{FFFFFF}91.93 &
  \multicolumn{1}{c|}{\textbf{98.99}} &
  25.67 &
  \textbf{8.45} \\
3 &
  \cellcolor[HTML]{FFFFFF} \textbf{94.09} &
  \multicolumn{1}{c|}{92.90} &
  \textbf{5.59} &
  11.88 \\
4 &
  \cellcolor[HTML]{FFFFFF}93.34 &
  \multicolumn{1}{c|}{\cellcolor[HTML]{FFFFFF}{\color[HTML]{212121} \textbf{97.56}}} &
  19.59 &
  \textbf{16.99} \\
5 &
  \cellcolor[HTML]{FFFFFF}89.32 &
  \multicolumn{1}{c|}{\cellcolor[HTML]{FFFFFF}{\color[HTML]{212121} \textbf{96.04}}} &
  \cellcolor[HTML]{FFFFFF}38.51 &
  \cellcolor[HTML]{FFFFFF} \textbf{24.10} \\
6 &
  \cellcolor[HTML]{FFFFFF}61.69 &
  \multicolumn{1}{c|}{\cellcolor[HTML]{FFFFFF} \textbf{88.91}} &
  \cellcolor[HTML]{FFFFFF}89.58 &
  \cellcolor[HTML]{FFFFFF} \textbf{24.34} \\
7 &
  \cellcolor[HTML]{FFFFFF}87.92 &
  \multicolumn{1}{c|}{\cellcolor[HTML]{FFFFFF} \textbf{97.01}} &
  \cellcolor[HTML]{FFFFFF}53.76 &
  \cellcolor[HTML]{FFFFFF} \textbf{24.43} \\
8 &
  \cellcolor[HTML]{FFFFFF}98.75 &
  \multicolumn{1}{c|}{\textbf{99.18}} &
  7.32 &
  \textbf{5.70} \\
9 &
  \cellcolor[HTML]{FFFFFF}85.68 &
  \multicolumn{1}{c|}{\textbf{98.80}} &
  \cellcolor[HTML]{FFFFFF}77.63 &
  \cellcolor[HTML]{FFFFFF} \textbf{17.21} \\
10 &
  \cellcolor[HTML]{FFFFFF} \textbf{96.91} &
  \multicolumn{1}{c|}{\cellcolor[HTML]{FFFFFF}93.19} &
  \cellcolor[HTML]{FFFFFF} \textbf{13.81} &
  \cellcolor[HTML]{FFFFFF}39.19 \\
11 &
  \cellcolor[HTML]{FFFFFF}81.40 &
  \multicolumn{1}{c|}{\cellcolor[HTML]{FFFFFF} \textbf{98.58}} &
  \cellcolor[HTML]{FFFFFF}41.75 &
  \cellcolor[HTML]{FFFFFF} \textbf{11.54} \\ \hline
\end{tabular}
}
\end{table}

 In Table \ref{foldwise_perform}, it can be noticed that apart from folds 3 and 10, our proposed MIBINET significantly outperforms the state-of-the-art HSMM in both metrics. To substantiate the robustness and efficacy of our proposed approach compared to its counterpart, we test their performance over various datasets of different modalities. These particular ones are chosen because of their availability and authenticity. In the following subsections, we present a brief account of these datasets.

\subsubsection{PCG Dataset}

Shi et al. \cite{shi2020dataset} also provide the phonocardiograph (PCG) signals corresponding to the Radar-HS data from the previously considered 11 test subjects. With the similar split, recording number, sampling rate, and ground truth of Radar-HS, it just differs in the signal modality. 

\subsubsection{PTB-XL ECG Dataset}
A 12-lead electrocardiography \cite{wagner2020ptb} dataset comprising of 21837 records from 18885 patients is used for comparison. The sampling rate is 100 Hz in the given recordings. However, we resample the dataset to 500 Hz to maintain conformity with the Radar-HS dataset. As it has a massive number of unique subjects under test, it can test the subject dependency of an algorithm. Hence, this dataset is used to demonstrate the universality of MIBINET. For dataset validity, it is split sequentially into train, validation, and test sets with a ratio of 60:20:20 based on the number of subjects. For the sake of simplicity, only the signal from lead II is considered, as the QRS complex here is more prominent compared to the ones in the other leads. It is to be noted that ECG signals are annotated using BioSPPy \cite{milivojevic2017python} as suggested in \cite{wagner2020ptb}. 

\begin{table}[]
\caption{Comparison between MIBINET and state-of-the-art HSMM algorithm}
\label{mibinetvshsmm}
\centering
\begin{tabular}{c|c|c|c}
\hline
Dataset                    & Methodology & $r$ (\%)  & RMSE (ms) \\ \hline \hline
\multirow{2}{*}{Radar-HS \cite{shi2020dataset}}   & HSMM       & 93.66 & 47.94   \\ \cline{2-4} 
& MIBINET     & \textbf{98.73} & \textbf{20.69}   \\ \hline
\multirow{2}{*}{PCG \cite{shi2020dataset}}       
& HSMM & 87.02 & 53.52   \\ \cline{2-4} 
& MIBINET   & \textbf{98.76} & \textbf{21.02}   \\ \hline
\multirow{2}{*}{PTB-XL ECG \cite{wagner2020ptb}} & HSMM         & 71.83 & 140.55   \\ \cline{2-4} 
& MIBINET  & \textbf{99.60} & \textbf{11.93}   \\ \hline
\multirow{2}{*}{Wrist PPG \cite{jarchi2016description}} & HSMM   & -5.55 & 2641.43   \\ \cline{2-4} 
& MIBINET    & \textbf{56.01} & \textbf{182.28} \\ \hline
\end{tabular}
\end{table}

\subsubsection{Wrist PPG}
Jarchi et al. \cite{jarchi2016description} released a PPG dataset collected using smartwatches from 8 participants. Here, all the recordings are sampled at a 256 Hz sampling rate, and 19 recordings are recorded in four different conditions, i.e., running, walking, fast-easy bike riding, and slow-difficult bike riding. We include this dataset as it has heavy motion artefacts, which is typical in real-life scenarios. Here, the R-peaks are determined from the given synchronized ECG signal. It also contains high HRV cases due to sudden initiation of motion and abrupt stoppages. Utilizing the 19 recordings, 19-fold cross-validation results are reported where fifteen, three, and one folds are in training, validation, and testing, respectively. 

Now, we present the cumulative performance of MIBINET and HSMM in the above four (Radar-HS, PCG, ECG, and PPG) datasets; see Table \ref{mibinetvshsmm}. Here, we observe the following:
\begin{enumerate}
    \item For the Radar-HS dataset, MIBINET provides an overall performance improvement of more than 5\% in terms of correlation and 27 ms in RMSE, compared to HSMM. 
    \item In the PCG dataset, our algorithm performed even better than in Radar-HS. A higher correlation coefficient and lower RMSE portrays this performance improvement.
    \item For the ECG dataset, the performance improvement of MIBINET over HSMM is even more prominent compared to the previous two datasets (correlation coefficient improvement $>$ 27\%).
    \item Although the PPG signal is more challenging due to its miscellaneous artefacts and high noise content, MIBINET still outperforms the HSMM by a massive margin. The results in the final row of Table \ref{mibinetvshsmm} show the robustness of our algorithm in challenging conditions induced by this PPG modality.
\end{enumerate}

\begin{figure*}

\captionsetup[subfloat]{farskip=5pt,captionskip=2pt}
\subfloat[]{
  \includegraphics[height=50mm,width=60mm]{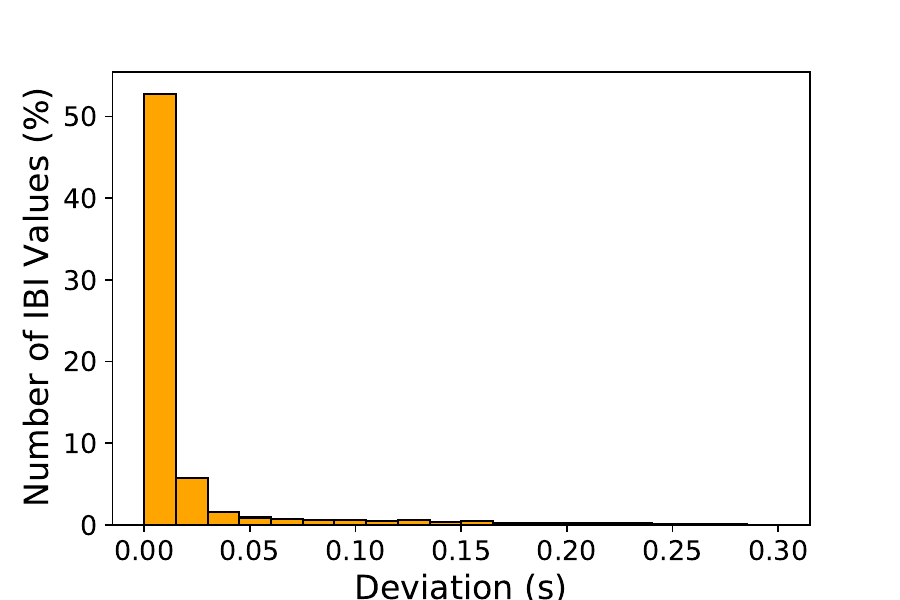}  
  \label{blandalt_a}
}
\subfloat[]{
  \includegraphics[height=50mm,width=59mm]{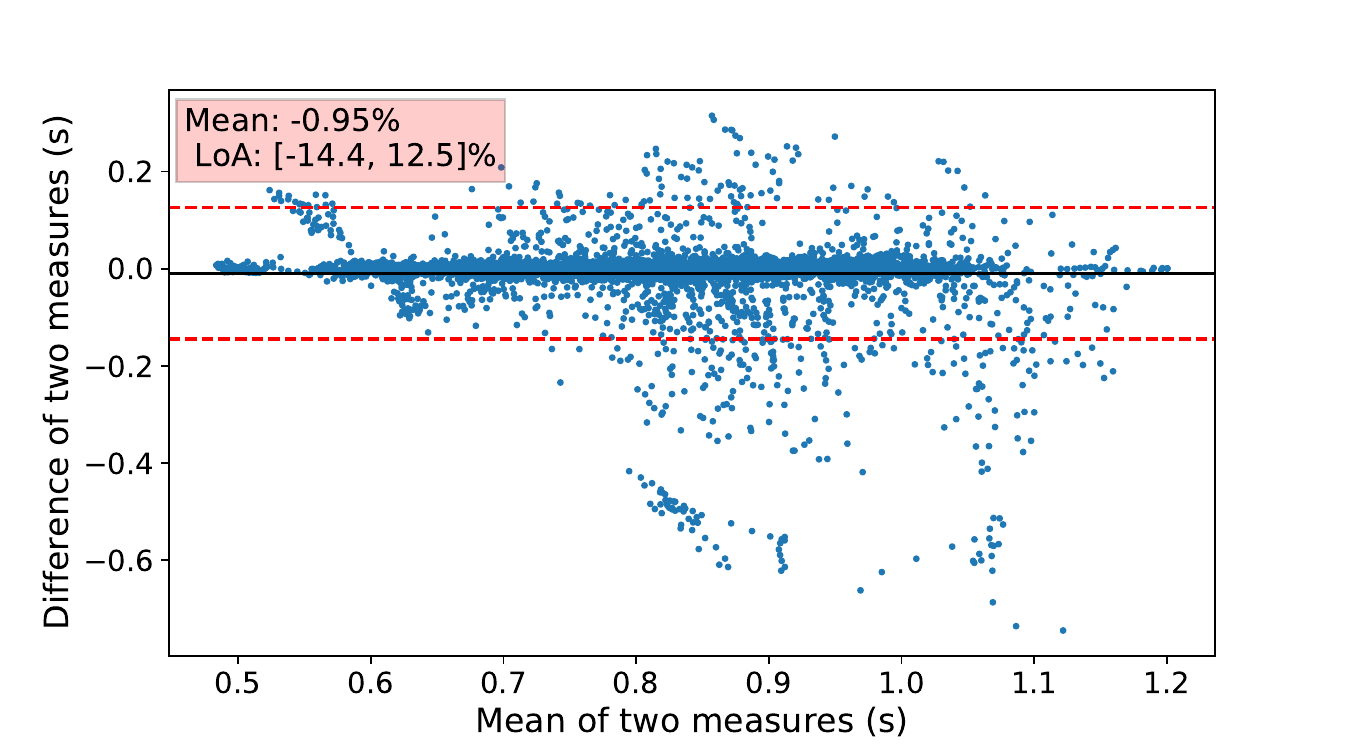}
  \label{blandalt_b}
}
\subfloat[]{
  \includegraphics[height=50mm,width=59mm]{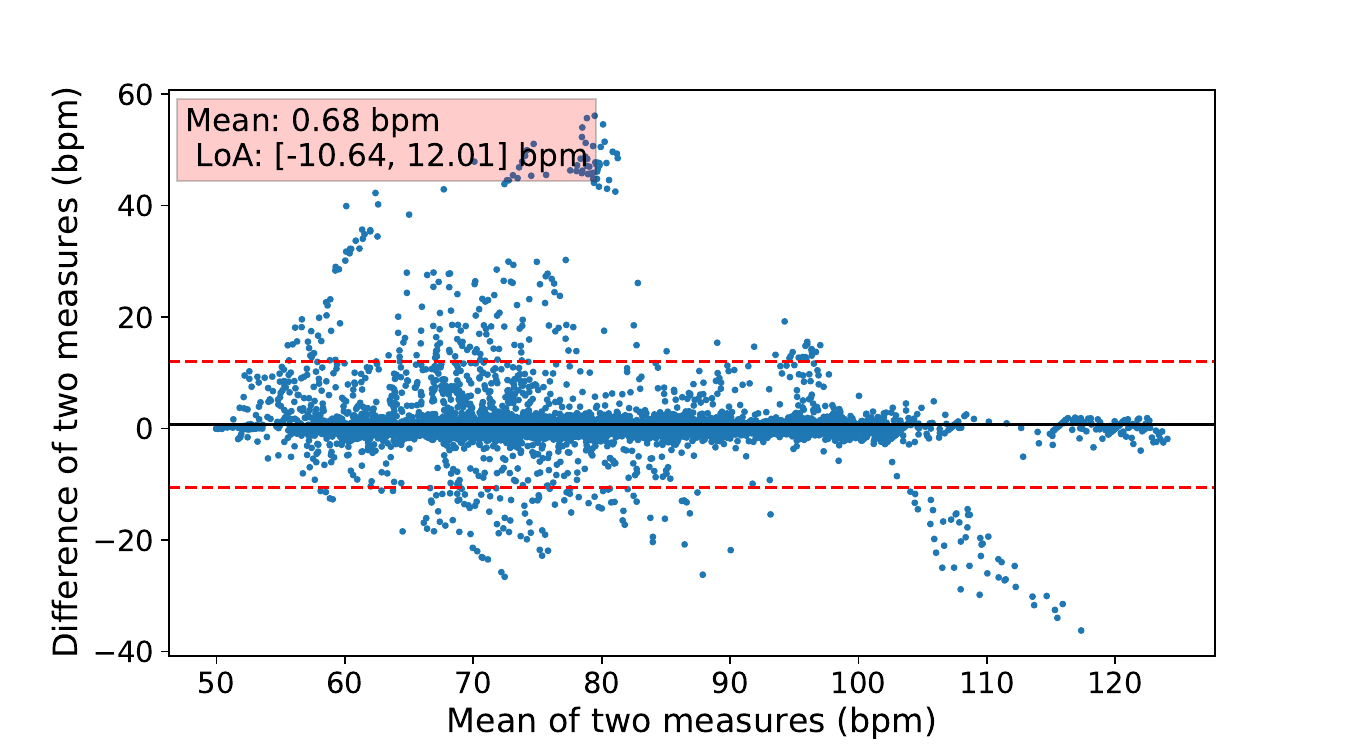}
  \label{blandalt_c}
}\\[-2ex]
\newline
\subfloat[]{
  \includegraphics[height=50mm,width=59mm]{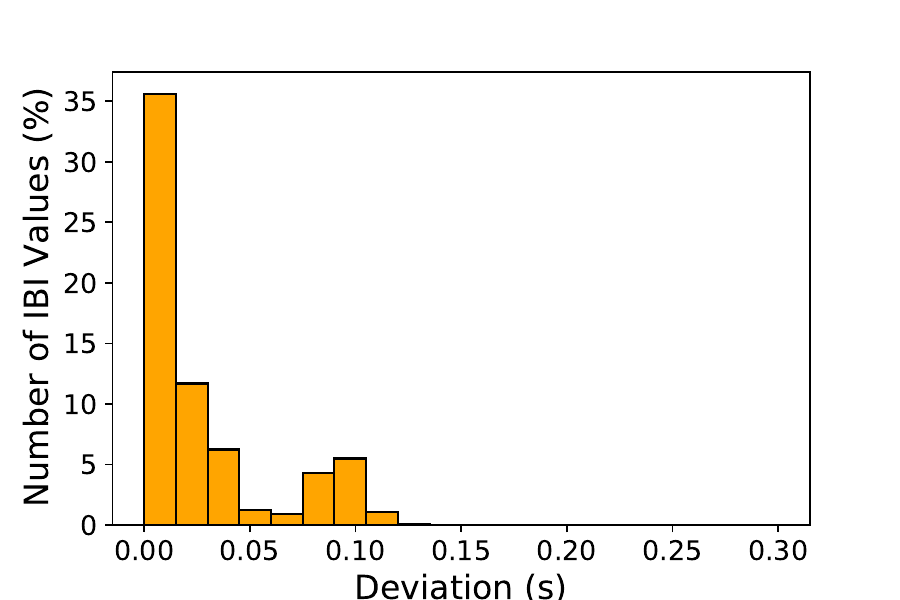}  
  \label{blandalt_d}
}
\subfloat[]{
  \includegraphics[height=50mm,width=59mm]{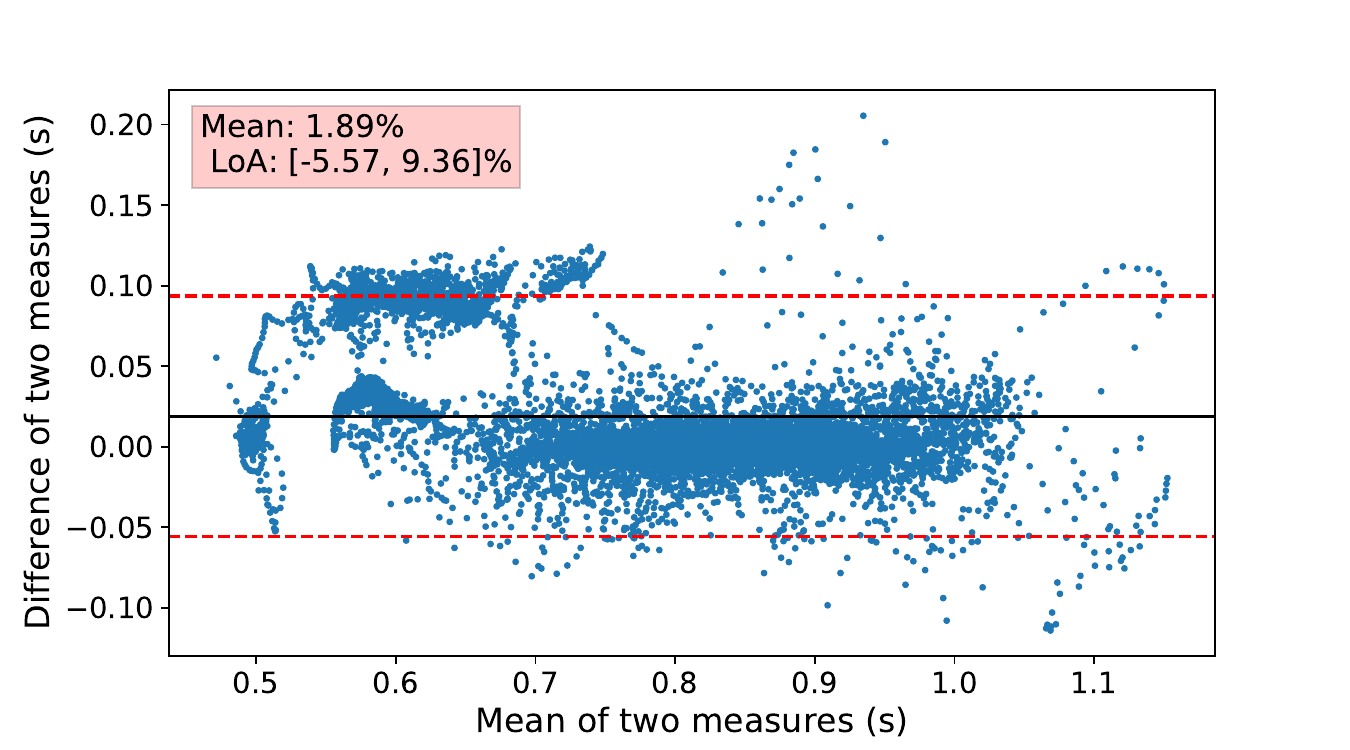}
  \label{blandalt_e}
}
\subfloat[]{
  \includegraphics[height=50mm,width=59mm]{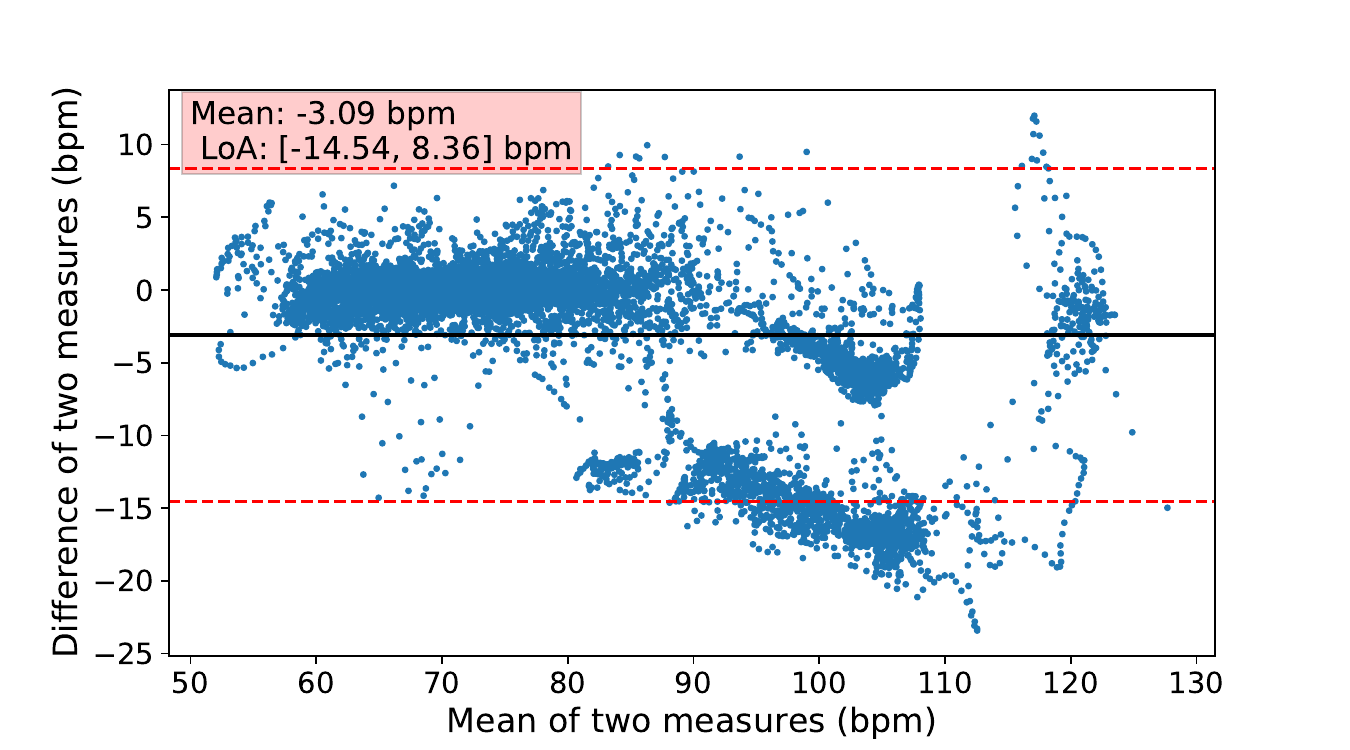}
  \label{blandalt_f}
}\\[-2ex]
\newline
\subfloat[]{
  \includegraphics[height=50mm,width=59mm]{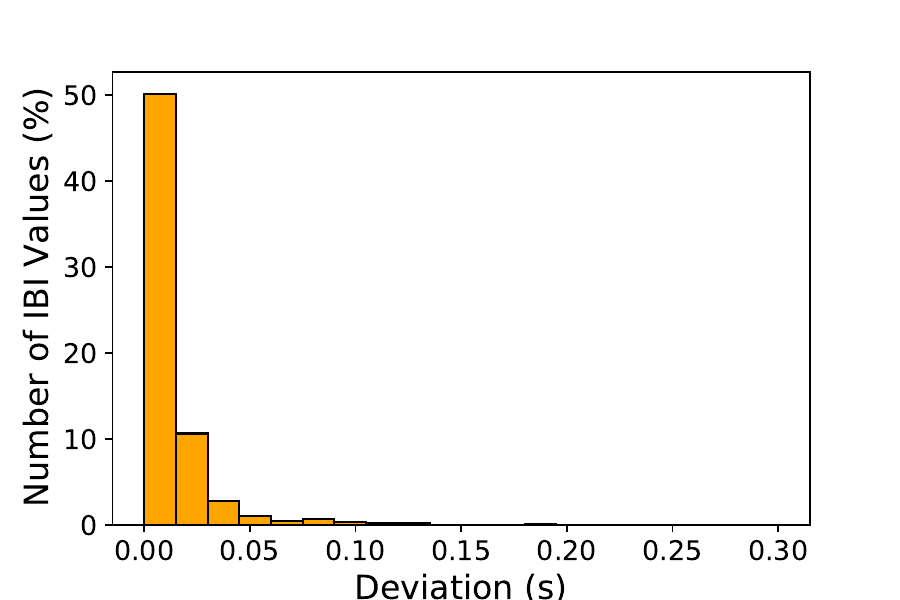}
  \label{blandalt_g}
}
\subfloat[]{
  \includegraphics[height=50mm,width=59mm]{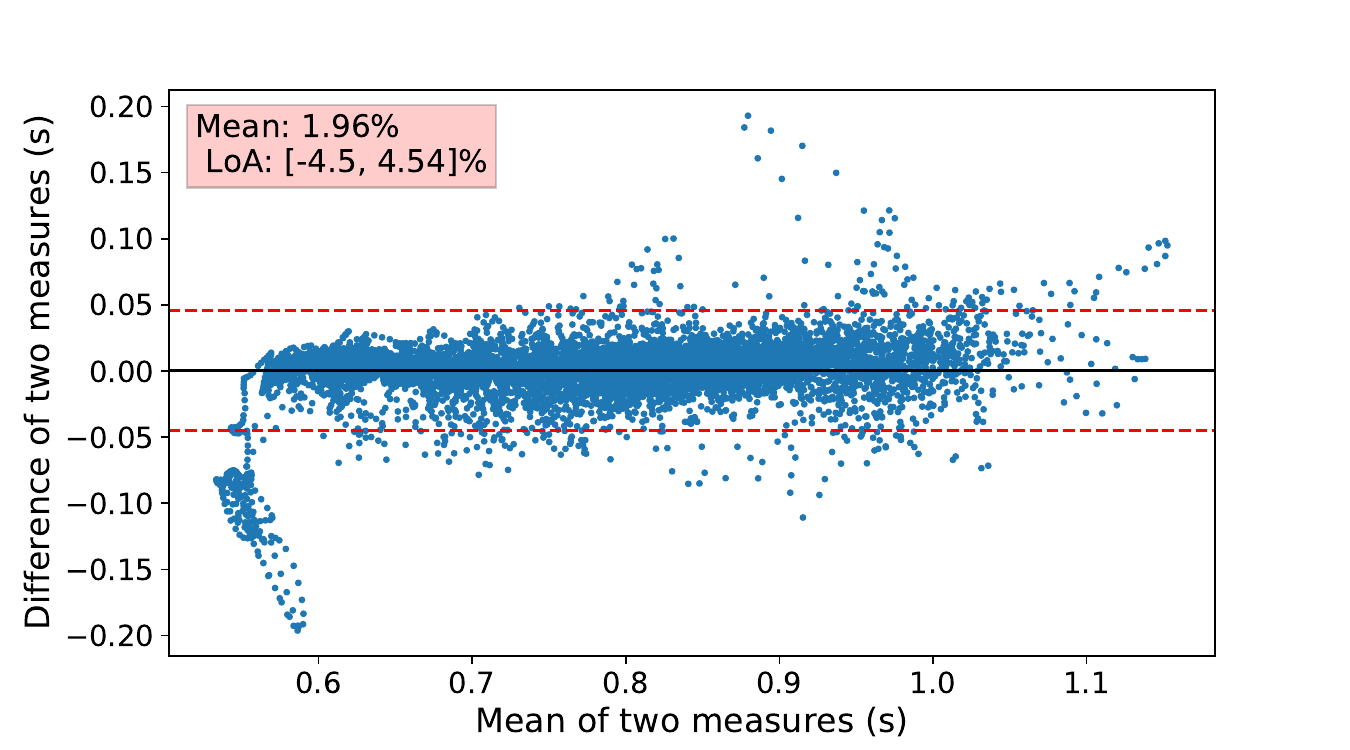}
  \label{blandalt_h}
}
\subfloat[]{
  \includegraphics[height=50mm,width=59mm]{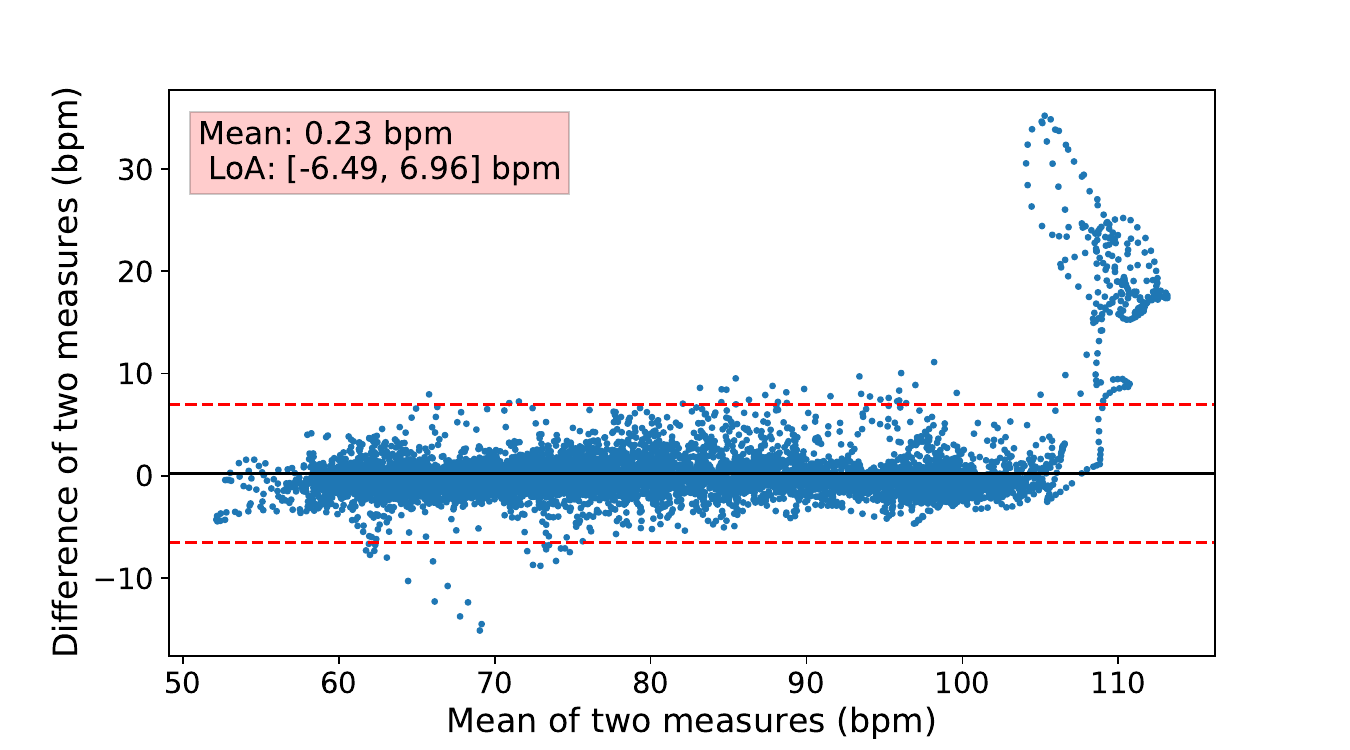}
  \label{blandalt_i}
}\\[-2ex]
\newline
\vspace{-2mm}
\subfloat[]{
  \includegraphics[height=50mm,width=59mm]{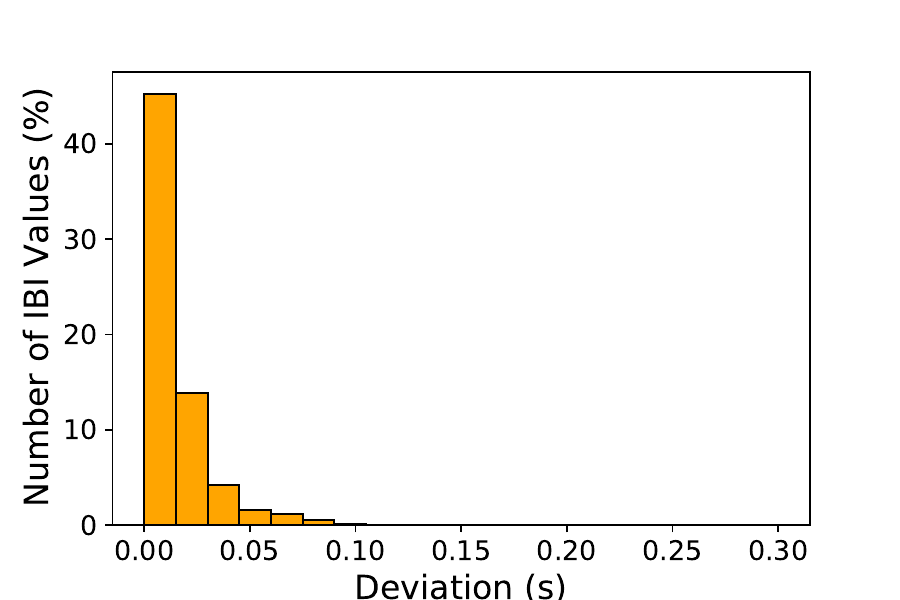}
  \label{blandalt_j}
}
\subfloat[]{
  \includegraphics[height=50mm,width=59mm]{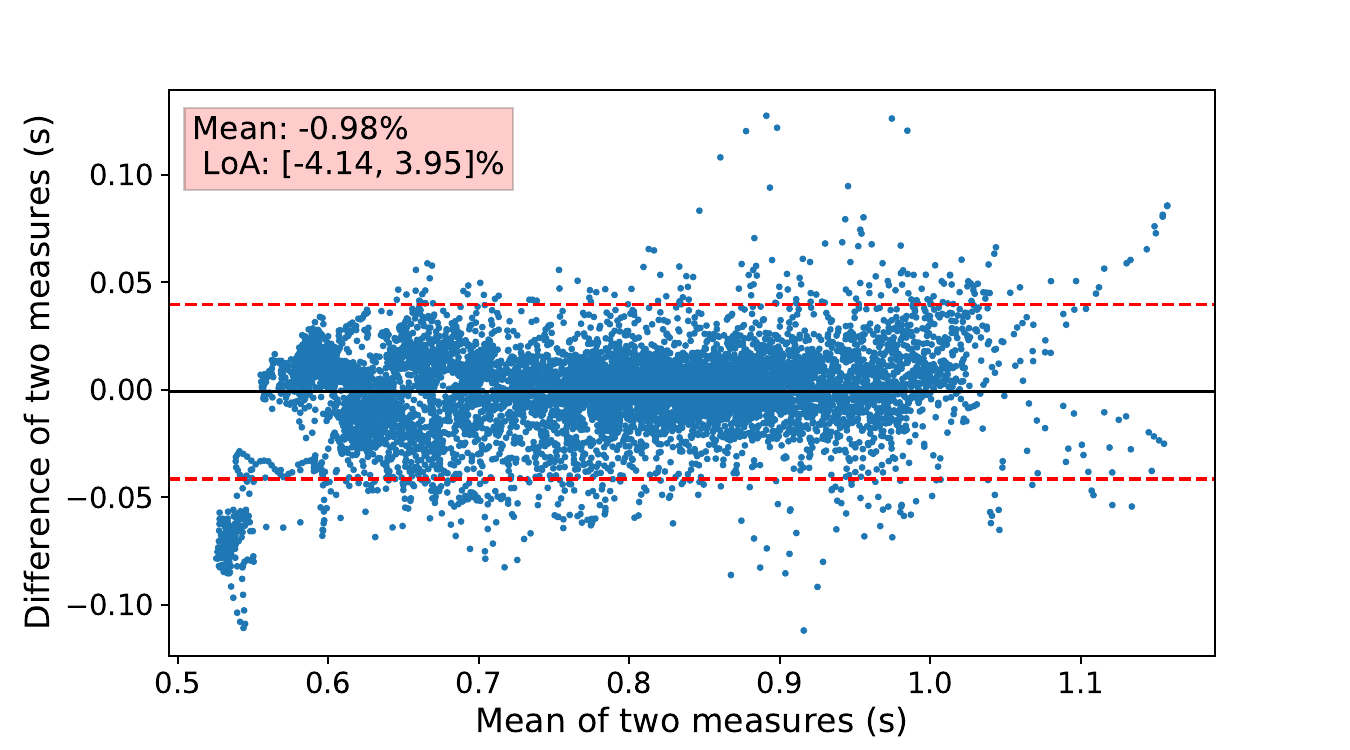}
  \label{blandalt_k}
}
\subfloat[]{
  \includegraphics[height=50mm,width=59mm]{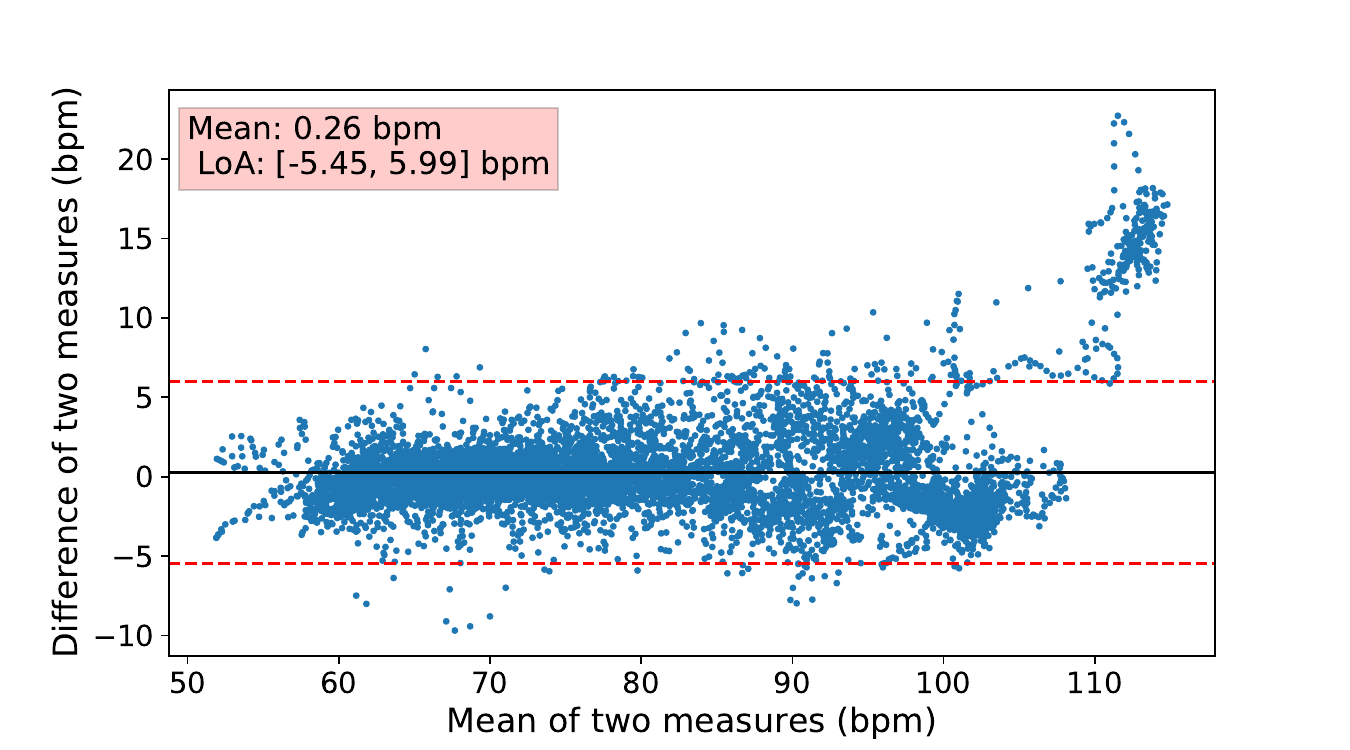}
  \label{blandalt_l}
}\\
\caption{Performance comparison based on error: (a) - (c) are plots for state-of-the-art HSMM-based algorithm.
(d) - (f) are plots for Huber loss with augmentation. (g) - (i) are plots for proposed loss without augmentation. (j) - (l) are for proposed loss with augmentation. In the first column, error deviation plots; in the second column, Bland Altman plots of deviation between ground truth (Radar-HS) and predicted values; and in the third column, Bland Altman plots of deviation in the BPM unit are depicted.}
\label{error_plot}
\end{figure*}

\subsection{Performance of Proposed Loss Function}


A major contribution of this work is the weighted custom loss introduced in (\ref{prop_loss}). The efficacy of the proposed loss is tested on various modalities of the dataset. We compare the performance of the proposed weighted loss with that of the best-performing existing regression loss, namely Huber-loss; see Table \ref{loss_table}. As can be observed, the proposed loss function assists MIBINET in improving its performance substantially in all datasets. This signifies that our proposed loss is not dataset-specific but performs equally well over other datasets. 

Next, we conduct Bland Altman analysis to show the agreement between predicted and ground truth IBI; see Figure \ref{error_plot}. In this analysis, we consider four different cases. The comparison between the true and predicted values is shown through the deviation plots and the Bland Altman plots of the IBI values and the corresponding BPM values. The plots in Figure \ref{error_plot} show the mean and standard deviation in radar heart sound data. The LoA (limits of agreement) is lower in our method than that of state-of-the-art, which is shown in Figure \ref{error_plot}(b - c) and \ref{error_plot}(k - l). The deviation plots also reveal a significant improvement over the HSMM method. We can clearly observe that there are some errors in the region of 0.1 s to 0.25 s for the HSMM deviation plot shown in \ref{error_plot}a. Figure \ref{error_plot}(d - f) vs. Figure \ref{error_plot}(j - l) shows the clear distinction between Huber loss and proposed loss, respectively. As we can observe, the regressor with Huber loss has many errors beyond 0.05 s in the deviation plot. Moreover, the LoA and the mean are better for the regressor with the proposed loss.


\begin{table}[]
\caption{Performance comparison of Proposed Loss}
\label{loss_table}
\centering
\resizebox{\columnwidth}{!}{
\begin{tabular}{c|c|c|c}
\hline
Dataset                    & Loss Function & $r$ (\%)   & RMSE (ms) \\ \hline \hline
\multirow{2}{*}{Radar-HS \cite{shi2020dataset}}  & Huber   & 97.31 & 42.55   \\ \cline{2-4} 
& Proposed   & \textbf{98.73} & \textbf{20.69} \\ \hline
\multirow{2}{*}{PCG \cite{shi2020dataset}}       & Huber   & 93.36 & 60.85   \\ \cline{2-4} 
& Proposed      & \textbf{98.76} & \textbf{21.02}   \\ \hline
\multirow{2}{*}{PTB-XL ECG \cite{wagner2020ptb}} & Huber    & 37.90 & 248.57   \\ \cline{2-4} 
& Proposed      & \textbf{99.60} & \textbf{11.93}   \\ \hline
\multirow{2}{*}{Wrist PPG \cite{jarchi2016description}} & Huber         & 52.56 & 1117.47   \\ \cline{2-4} 
& Proposed      & \textbf{56.01} & \textbf{182.28}   \\ \hline
\end{tabular}
}
\end{table}


\subsection{Effectiveness of Synthetic IBI Generation}

We again perform the Bland-Altman analysis to demonstrate the potency of our synthetic IBI generation process. It can be clearly noticed from Figure \ref{error_plot}(h - i) and \ref{error_plot}(k - l) that, our proposed augmentation method decreases the error and gives a better LoA. In purely quantitative terms, the RMSE is decreased by almost 3 ms for the Radar-HS dataset. Similarly, the results obtained through PCG data also show the effectiveness of this novel synthetic IBI augmentation. In PCG modality, we observe a performance improvement of around 10\% in correlation coefficient and 40 ms in RMSE, as displayed in Table \ref{aug_perform}.


\begin{table}[]
\caption{Performance comparison of augmentation on MIBINET}
\label{aug_perform}
\centering
\begin{tabular}{c|c|c|c}
\hline
Dataset                    & Augmentation & $r$ (\%)  & RMSE (ms) \\ \hline \hline
\multirow{2}{*}{Radar-HS \cite{shi2020dataset}}   & -       & 98.44 & 23.08   \\ \cline{2-4} 
& \checkmark   & \textbf{98.73}
& \textbf{20.69}   \\ \hline
\multirow{2}{*}{PCG \cite{shi2020dataset}}   & -       & 88.46 & 61.03  \\ \cline{2-4} 
& \checkmark   & \textbf{98.77}
& \textbf{21.02}   \\ \hline
\end{tabular}
\end{table}

\vspace{-1mm}

\subsection{Comparison with Related Models}

Besides HSMM, we compare the performance of the MIBINET model against the current state-of-the-art lightweight models applicable to the Radar-HS dataset provided in \cite{shi2020dataset}. To evaluate the performance of the neural network models alone, we test them without post-processing, and the results are presented in Table \ref{comparison}. It is to be noted that all the peripheral settings were kept exactly the same for all the models to ensure a fair comparison. We notice that MIBINET outperforms the current state-of-the-art lightweight deep learning models by a margin of at least 3\% in the correlation coefficient metric. Importantly, MIBINET, having the least number of parameters among all these models, claims itself as the best-suited real-time solution for contactless vital signs monitoring.

\begin{table}[h]
\centering
\caption{Comparison with lightweight Neural Networks on the hold-out test set of Radar-HS (before post-processing filtering)}
\label{comparison}
\resizebox{\linewidth}{!}{
\begin{tabular}{lcc}
\hline 
Model Name    & $r$ (\%)  & Parameters ($\times 10^6$)\\ 
\hline \hline
MobileNet V1 \cite{howard2017mobilenets} & $84.57$ & $7.98$ \\
MobileNet V2 \cite{sandler2018mobilenetv2} & $84.55$ & $10.93$ \\
MobileNet V3 small \cite{howard2019searching} & $84.44$ & $1.28$ \\
MobileNet V3 large \cite{howard2019searching} & $84.42$ & $2.99$ \\
ResNet 34 \cite{he2016deep} & $78.06$ & $7.07$ \\
ResNet 50 \cite{he2016deep} & $85.66$ & $23.74$ \\
MIBINET & \textbf{88.57} & \textbf{1.09} \\
\hline
\end{tabular}
}
\end{table}

\section{Discussion}
\label{discuss}
This work proposed a comprehensive DNN-based strategy (MIBINET) to surpass the present state-of-the-art (HSMM) in measuring IBI from mm-wave collected data. A key aspect of our work lies in the integration of novel signal processing schemes with the excellent feature extraction capabilities of CNN. MIBINET is a unique lightweight architecture that features innovative post-processing techniques, including rolling window averaging and traditional filtering, which contribute to the improved performance of the model. The proposed approach achieved a cumulative 98.73\% correlation coefficient and produced a meagre 20.69 ms RMSE score over 11 different test subjects on Radar-HS data. This was a significant improvement over HSMM, which offered a 93.66\% correlation coefficient and 47.94 ms RMSE.

The robustness and versatility of our proposed approach were further demonstrated by evaluating it on datasets containing ECG, PCG, and PPG signals, where its performance was found to be superior to HSMM in all of them. A custom-weighted regression loss was developed to train MIBINET, and a unique weight-saving approach was used based on a weighted metric. The novel weighted loss function improved the correlation coefficient by more than 1.4\% and lowered the RMSE by more than 21 ms in Radar-HS data compared to the commonly used Huber-loss function.

A novel IBI augmentation technique was also introduced to negate the effects of dataset imbalance, which led to significant performance improvements. However, there are some limitations and future research directions, such as the applicability of the synthetic IBI augmentation technique on non-stationary signals like ECG and PPG and the performance of MIBINET in scenarios with high HRV.

\section{Conclusion}
\label{conclusion}
In conclusion, our work has demonstrated the effectiveness of integrating novel signal processing techniques with the feature extraction capabilities of CNN in the context of estimating IBI values from various signal modalities such as Radar-HS, PCG, ECG, and PPG. The proposed MIBINET approach showcased significant improvements over the current state-of-the-art methods. Our findings also highlighted the versatility and robustness of 
multifaceted MIBINET across diverse signal modalities, as its performance consistently surpassed that of the HSMM method. Moreover, our custom-weighted regression loss and the novel IBI augmentation technique effectively addressed dataset imbalance, leading to substantial performance enhancements. These results underline the potential of MIBINET in contactless vital signs monitoring, offering a promising solution for various real-life applications in medical, home, and transportation settings. Additionally, we believe that integrating our newly designed signal processing techniques can significantly enhance the performance of other 1D models.

Looking forward, there are several avenues for future research to further enhance and expand upon the capabilities of MIBINET. These include exploring the use of raw radar data for end-to-end modelling, addressing limitations in synthetic IBI augmentation for ECG and PPG data, and considering multi-user scenarios to make the model more practicable. Additionally, investigating multi-modal approaches that combine multiple signals or fuse multiple streams instead of relying on single-channel data could open up new dimensions in the field of contactless vital signs monitoring.

\section*{Conflict of interest}The authors declare no conflict of interest.

\balance
\bibliography{main.bib} 

\end{document}